\documentclass[11pt]{article}
\usepackage{amsfonts}
\usepackage{amssymb}
\usepackage{amsmath}
\usepackage{amsxtra}
\usepackage{graphicx}
\usepackage{psfrag}
\usepackage{mathrsfs}
\usepackage{natbib}
\usepackage{stmaryrd}
\usepackage{subfig}
\usepackage{tikz}
\usepackage{empheq}
\usepackage{pgfplots}
\usepackage{pgfplotstable}
\pgfplotsset{compat=1.3}
\pgfplotsset{every axis legend/.append style={
    at={(1.05,1)},
    anchor=north west,font=\small}}
\usepackage{filecontents}
\usepackage{float}
\usepgfplotslibrary{units}

\begin{document}
\def\BGamma{\mbox{\boldmath$\Gamma$}}
\def\BDelta{\mbox{\boldmath$\Delta$}}
\def\BTheta{\mbox{\boldmath$\Theta$}}
\def\BLambda{\mbox{\boldmath$\Lambda$}}
\def\BXi{\mbox{\boldmath$\Xi$}}
\def\BPi{\mbox{\boldmath$\Pi$}}
\def\BSigma{\mbox{\boldmath$\Sigma$}}
\def\BUpsilon{\mbox{\boldmath$\Upsilon$}}
\def\BPhi{\mbox{\boldmath$\Phi$}}
\def\BPsi{\mbox{\boldmath$\Psi$}}
\def\BOmega{\mbox{\boldmath$\Omega$}}
\def\Balpha{\mbox{\boldmath$\alpha$}}
\def\Bbeta{\mbox{\boldmath$\beta$}}
\def\Bgamma{\mbox{\boldmath$\gamma$}}
\def\Bdelta{\mbox{\boldmath$\delta$}}
\def\Bepsilon{\mbox{\boldmath$\epsilon$}}
\def\Bzeta{\mbox{\boldmath$\zeta$}}
\def\Beta{\mbox{\boldmath$\eta$}}
\def\Btheta{\mbox{\boldmath$\theta$}}
\def\Biota{\mbox{\boldmath$\iota$}}
\def\Bkappa{\mbox{\boldmath$\kappa$}}
\def\Blambda{\mbox{\boldmath$\lambda$}}
\def\Bmu{\mbox{\boldmath$\mu$}}
\def\Bnu{\mbox{\boldmath$\nu$}}
\def\Bxi{\mbox{\boldmath$\xi$}}
\def\Bpi{\mbox{\boldmath$\pi$}}
\def\Brho{\mbox{\boldmath$\rho$}}
\def\Bsigma{\mbox{\boldmath$\sigma$}}
\def\Btau{\mbox{\boldmath$\tau$}}
\def\Bupsilon{\mbox{\boldmath$\upsilon$}}
\def\Bphi{\mbox{\boldmath$\phi$}}
\def\Bchi{\mbox{\boldmath$\chi$}}
\def\Bpsi{\mbox{\boldmath$\psi$}}
\def\Bomega{\mbox{\boldmath$\omega$}}
\def\Bvarepsilon{\mbox{\boldmath$\varepsilon$}}
\def\Bvartheta{\mbox{\boldmath$\vartheta$}}
\def\Bvarpi{\mbox{\boldmath$\varpi$}}
\def\Bvarrho{\mbox{\boldmath$\varrho$}}
\def\Bvarsigma{\mbox{\boldmath$\varsigma$}}
\def\Bvarphi{\mbox{\boldmath$\varphi$}}
\def\bone{\mbox{\boldmath$1$}}
\def\bzero{\mbox{\boldmath$0$}}
\def\bA{\mbox{\boldmath$ A$}}
\def\bB{\mbox{\boldmath$ B$}}
\def\bC{\mbox{\boldmath$ C$}}
\def\bD{\mbox{\boldmath$ D$}}
\def\bE{\mbox{\boldmath$ E$}}
\def\bF{\mbox{\boldmath$ F$}}
\def\bG{\mbox{\boldmath$ G$}}
\def\bH{\mbox{\boldmath$ H$}}
\def\bI{\mbox{\boldmath$ I$}}
\def\bJ{\mbox{\boldmath$ J$}}
\def\bK{\mbox{\boldmath$ K$}}
\def\bL{\mbox{\boldmath$ L$}}
\def\bM{\mbox{\boldmath$ M$}}
\def\bN{\mbox{\boldmath$ N$}}
\def\bO{\mbox{\boldmath$ O$}}
\def\bP{\mbox{\boldmath$ P$}}
\def\bQ{\mbox{\boldmath$ Q$}}
\def\bR{\mbox{\boldmath$ R$}}
\def\bS{\mbox{\boldmath$ S$}}
\def\bT{\mbox{\boldmath$ T$}}
\def\bU{\mbox{\boldmath$ U$}}
\def\bV{\mbox{\boldmath$ V$}}
\def\bW{\mbox{\boldmath$ W$}}
\def\bX{\mbox{\boldmath$ X$}}
\def\bY{\mbox{\boldmath$ Y$}}
\def\bZ{\mbox{\boldmath$ Z$}}
\def\ba{\mbox{\boldmath$ a$}}
\def\bb{\mbox{\boldmath$ b$}}
\def\bc{\mbox{\boldmath$ c$}}
\def\bd{\mbox{\boldmath$ d$}}
\def\be{\mbox{\boldmath$ e$}}
\def\bff{\mbox{\boldmath$ f$}}
\def\bg{\mbox{\boldmath$ g$}}
\def\bh{\mbox{\boldmath$ h$}}
\def\bi{\mbox{\boldmath$ i$}}
\def\bj{\mbox{\boldmath$ j$}}
\def\bk{\mbox{\boldmath$ k$}}
\def\bl{\mbox{\boldmath$ l$}}
\def\bm{\mbox{\boldmath$ m$}}
\def\bn{\mbox{\boldmath$ n$}}
\def\bo{\mbox{\boldmath$ o$}}
\def\bp{\mbox{\boldmath$ p$}}
\def\bq{\mbox{\boldmath$ q$}}
\def\br{\mbox{\boldmath$ r$}}
\def\bs{\mbox{\boldmath$ s$}}
\def\bt{\mbox{\boldmath$ t$}}
\def\bu{\mbox{\boldmath$ u$}}
\def\bv{\mbox{\boldmath$ v$}}
\def\bw{\mbox{\boldmath$ w$}}
\def\bx{\mbox{\boldmath$ x$}}
\def\by{\mbox{\boldmath$ y$}}
\def\bz{\mbox{\boldmath$ z$}}
\def\bdX{\mbox{\boldmath$ dX$}}
\def\bdx{\mbox{\boldmath$ dx$}}
\newcommand*\mycirc[1]{%
  \begin{tikzpicture}
    \node[draw,circle,inner sep=1pt] {#1};
  \end{tikzpicture}
}

\title{Three-dimensional iso-geometric solutions to general boundary value problems of Toupin's gradient elasticity theory at finite strains}
\date{}
\author{S.~Rudraraju\thanks{Department of Mechanical Engineering, University of Michigan, Ann Arbor, {\tt rudraa@umich.edu}}, A.~van~der~Ven\thanks{Materials Department, University of California at Santa Barbara, {\tt avdv@engineering.ucsb.edu}} \& K.~Garikipati\thanks{Departments of Mechanical Engineering and Mathematics, University of Michigan, Ann Arbor, corresponding author, {\tt krishna@umich.edu}}}
\maketitle
\abstract{We present, to the best of our knowledge, the first complete three-dimensional solutions to a broad range of boundary value problems for a general theory of finite strain gradient elasticity. We have chosen for our work, Toupin's theory [\emph{Arch. Rat. Mech. Anal.}, \textbf{11}(1), 385-414, 1962]--one of the more general formulations of strain gradient elasticity. Our framework has three crucial ingredients: The first is iso-geometric analysis [Hughes et al., \emph{Comp. Meth. App. Mech. Engrg.}, \textbf{194}(39-41), 4135-4195, 2005], which we have adopted for its straightforward and robust representation of $C^1$-continuity. The second is a weak treatment of the higher-order Dirichlet boundary conditions in the formulation, which control the development of strain gradients in the solution. The third ingredient is algorithmic (automatic) differentiation, which eliminates the need for linearization ``by hand'' of the rather complicated geometric and material nonlinearities in gradient elasticity at finite strains. We present a number of numerical solutions to demonstrate that the framework is applicable to arbitrary boundary value problems in three dimensions. We discuss size effects, the role of higher-order boundary conditions, and perhaps most importantly, the relevance of the framework to problems with elastic free energy density functions that are non-convex in strain space.}

%
%
\section{Introduction}
\label{sec:introduction} 
A fundamental assumption of the classical theory of elasticity is that the elastic free energy density is a frame-invariant function of the deformation gradient, $W = \widehat{W}(\bF)$ \citep{Truesdell1962}, where the deformation gradient is $\bF = \bone + \partial\bu/\partial\bX$, with $\bu$ being the displacement and $\bX$ being the reference placement.
\begin{figure}[h]
  \centering
    \psfrag{a}{\small $\Omega_0$}
    \psfrag{b}{$\Omega$}
    \psfrag{c}{$\bX$}
    \psfrag{d}{\small $\bX+\mathrm{d}\bX$}
    \psfrag{e}{$\mathrm{d}\bX$}
    \psfrag{f}{$\bx$}
    \psfrag{g}{\small $\bx+\mathrm{d}\bx$}
    \psfrag{h}{$\mathrm{d}\bx$}
    \psfrag{i}{$\Bvarphi(\bX)$}
    \psfrag{o}{$O$}
  \includegraphics[width=0.4\textwidth]{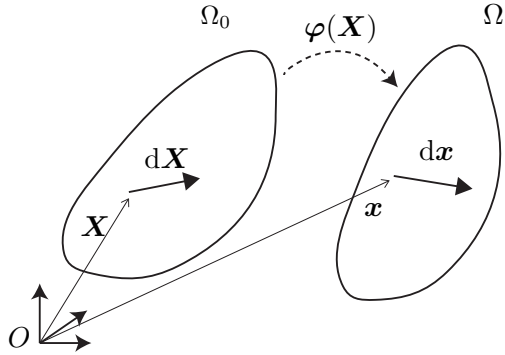}
\caption{Deformation of a line element $\mathrm{d}\bX$ in the reference configuration $\Omega_0$ to a line element $\mathrm{d}\bx$ in the current configuration under the deformation map $\Bvarphi(\bX)$.}
\label{fig:deformation}
\end{figure}

The connection between this continuum model and the atomic structure is that elastic free energy is stored due to bond stretching, which in turn is determined by the deformation field in an arbitrarily small neighborhood of a point $\bX$. A line element $\mathrm{d}\bX$ in the reference configuration $\Omega_0$ undergoes stretch and rotation to $\mathrm{d}\bx$ in the deformed configuration $\Omega$, as shown in Figure \ref{fig:deformation}. Using coordinate notation for clarity, the Taylor series expansion gives
\begin{equation}
\mathrm{d}x_i =  F_{iJ} \mathrm{d}X_J + F_{iJ,K} \mathrm{d}X_J \mathrm{d}X_K + F_{iJ,KL} \mathrm{d}X_J \mathrm{d}X_K \mathrm{d}X_L + \cdots
\label{eq:deformationMap}
\end{equation}
where $F_{iJ}=\delta_{iJ}+\partial u_i/\partial X_J$ is the coordinate representation of the deformation gradient tensor.  When the above expansion is extended down to the atomic scale, $\mathrm{d}\bx$ represents the change in bond length, which evidently depends on $\bF$ and its gradients. Therefore, the elastic free energy density parametrized by continuum fields also sports such a dependence. As shown by \citet{Garikipati2003}, if the elastic free energy depends on bond angle in addition to bond stretch, there is a further dependence on the first gradient of $\bF$. These dependencies, however, are strong only if the continuum deformation field varies sharply in all directions over the length scale of an atomic bond. This is not the case in traditional applications of elasticity, and the higher-order terms involving gradients of $\bF$ in Equation \eqref{eq:deformationMap} can be neglected. 

Introducing the Green-Lagrange strain, $\bE = \frac{1}{2}(\bF^\mathrm{T}\bF-\bone)$,
to ensure frame invariance, the classical elastic free energy density depends only on $\bE$, i.e. $W = \widetilde{W}(\bE)$. The gradients of $\bF$, or to ensure frame invariance, gradients of $\bE$, become relevant under two situations: (a) if strains also serve as order parameters in representations of symmetry-lowering structural phase transformations, leading to a non-convex elastic free energy density, and (b)  if variations in deformation occur over length scales approaching inter-atomic distances, as at atomically sharp crack tips and dislocation cores. In such cases the elastic free energy density must be extended to include strain gradients as $W = \widehat{W}(\bF,\mathrm{Grad}\bF) = \widetilde{W}(\bE, \mathrm{Grad}\bE)$. Dimensional analysis reveals that the inclusion of $\textrm{Grad}\bE$ introduces length scales in $\widetilde{W}(\bE,\textrm{Grad}\bE)$ relative to the coefficients of $\bE$, thereby eliminating the scale invariance of classical elasticity. We will postpone consideration of these two cases until the end of this communication, because they deserve a detailed discussion that would distract from our main goal here. 

The foundations of strain gradient elasticity were laid down in the 1960s [see \cite{Toupin1962, Toupin1964, Mindlin1964, Mindlin1965, Koiter1964}]. More than a century ago, however, the problem of generalized continua had been addressed by the brothers, Cosserat \citep{Cosserats1909}, and was elaborated upon much later by \cite{Eringen1976}. Further development and application of strain gradient elasticity has focussed on modeling fracture \citep{Sternberg1967}, and resolving the dislocation core \citep{Lazaretal2006} as already mentioned. 

Because of the introduction of a length scale related to the strain gradients, the extension of these theories to inelasticity has allowed the numerical treatment of strain localization in softening continua free of the well-known mesh-dependent pathology that otherwise plagues such computations \citep{Triantafyllidis1986, Aifantis1992, Altan1997, Wellsetal2004, Molarietal2006}. It also has fueled an extensive literature in strain gradient plasticity \citep{Fleck1997, Gao1999, Muhlhaus1991, DeBorst1992, Acharya2000, Gurtin2000, Gurtin2005}. These strain gradient theories for inelastic materials are beyond the scope of this communication. 

It is important to note that strain gradient formulations, whether for elasticity or inelasticity, admit analytical solutions of boundary value problems only in the simplest of cases. Standard finite element methods also are ill-suited to solving strain gradient (in)elasticity problems as we now explain: Returning the focus to strain gradient elasticity, we note that, when derived in a consistent manner from variational principles, the strong form of the problem is of fourth order. It is equivalent to a weak form that, in the linear setting, naturally induces the $\mathscr{H}^2$-norm on trial solutions and weighting functions. The corresponding finite-dimensional approximations are guaranteed to lie in $\mathscr{H}^2$ if drawn from a function space that is at least $C^1$-continuous. Classical finite element function spaces are only $C^0$-continuous across element interfaces and therefore are a poor choice. Extensions of the classical finite element formulations to enforce strong $C^1$ continuity have been developed using Hermite elements \citep{Papanicolopulos2009}, and to enforce weak $C^1$ continuity using discontinuous Galerkin methods \citep{Engeletal2002, Wellsetal2004, Molarietal2006, Maraldi2011},  and mixed formulations \citep{Zybell2012}. However, Hermite elements are not considered competitive for higher dimensions. The drawback of discontinuous Galerkin formulations is that they come with stability requirements that can prove challenging to impose, while forcing a proliferation of degrees of freedom. Finally, our experience in the course of the current work has been that mixed formulations are limited by the complicated boundary conditions of higher order that result for finite strain gradient elasticity formulations. \\
 
 This work presents a broad numerical framework for solving general boundary value problems in finite strain gradient elasticity. We adopt Toupin's formulation of gradient elasticity at finite strains \citep{Toupin1962} as it is one of the more general of such theories. As indicated above, the first numerical obstacle to overcome is the requirement of a $C^1$-continuous basis. For this purpose, we have adopted Isogeometric Analysis (IGA) \citep{HughesCottrellBazilevs2005, CottrellHughesBazilevs2009}, a mesh-based numerical method with Non-Uniform Rational B-Splines (NURBS) basis functions. The NURBS basis enables the construction of $C^n$-continuous function spaces, and is thus naturally suited for solving higher order partial differential equations. Our use of the spline basis for strain gradient elasticity is not a first: See \cite{Fischeretal2011} for linearized, infinitesimal strain elasticity, and \cite{Fischeretal20102, Fischeretal2010} for finite strain elasticity in this regard. While these papers consider two dimensional problems our framework is three-dimensional; we also have addressed the full generality of boundary conditions that arise in Toupin's theory. Of crucial importance is the enforcement of higher-order boundary conditions that arise in the variational setting, which is the point of departure for our work. This point cannot be overemphasized: For certain boundary value problems, the higher-order boundary conditions dominate the solution, even trumping the effect of the strain gradient length scale parameter. Their imposition in the Galerkin weak form is the second crucial ingredient in our framework. The third is the exact linearization of the Galerkin weak form--a tall task if attempted manually, given the extent of nonlinearity that is induced by Toupin's theory. Instead, we have turned to algorithmic differentiation using the Sacado library \citep{Sacado2005,sacado2012}.
 
 The main body of the paper begins with a summary of Toupin's theory. It then proceeds through a section detailing the above numerical methods, and another with several numerical examples that demonstrate the generality of the framework in application to three-dimensional boundary value problems. The paper concludes with a discussion, where we also suggest the potential for applications of our methods.
  
\section{Variational formulation}
\label{sec:math}
We follow a variational approach to arrive at the steady state equilibrium equations of finite strain gradient elasticity. Our treatment is posed in the Cartesian coordinate system, with basis vectors $\be_i$, $i = 1,\dots 3$, $\be_i\cdot\be_j = \delta_{ij}$. The reference configuration, its boundary and the surface normal at any boundary point are denoted by $\Omega_0$, $\partial \Omega_0$ and $\bN$, respectively, with $\vert\bN\vert = 1$. The corresponding entities in the current configuration are denoted by $\Omega$, $\partial \Omega$ and $\bn$, respectively. We work mostly with coordinate notation. Upper case indices are used to denote the components of vectors and tensors in the reference configuration and lower case indices are reserved for those in the current configuration. In this section, the variational formulation is presented on the reference configuration. The corresponding derivation in the current configuration appears in the Appendix. In the reference configuration, we consider the boundary to be the union of a finite number of smooth surfaces $\Gamma_0$, smooth edges $\Upsilon_0$ and corners $\Xi_0$: $\partial \Omega_0 = \Gamma_0 \cup \Upsilon_0 \cup \Xi_0$.  For functions defined on $\partial \Omega_0$, when necessary, the gradient operator is decomposed into the normal gradient operator $D$ and the surface gradient operator $D_{K}$, 
\begin{align}
\psi_{,K} &= D \psi N_{K} + D_{K} \psi\nonumber\\
\textrm{where}\quad D \psi N_{K} &= \psi_{,I} N_{I}N_{K}\;\textrm{and}\; D_{K} \psi = \psi_{,K} - \psi_{,I} N_{I}N_{K}
\label{surfacenormalgradient}
\end{align}

\noindent A material point is denoted by $\bX\, \in\, \Omega_0$. The deformation map between $\Omega_0$ and $\Omega$ is given by $\Bvarphi(\bX,t) =  \bX + \bu = \bx$, where $\bu$ is the displacement field. The deformation gradient is $\bF = \partial\Bvarphi/\partial\bX = \bone + \partial\bu/\partial\bX$, which in coordinate notation has already been expressed as $F_{iJ} = \partial \varphi_{i} / \partial X_{J} = \delta_{iJ} + \partial u_{i}/\partial X_{J}$. The Green-Lagrange strain tensor in coordinate notation is given by $E_{IJ} = \frac{1}{2}(F_{kI}F_{kJ} - \delta_{IJ})$. The Gibbs potential of the system is given by the following functional defined over the reference configuration:\footnote{When we refer to the \emph{elastic free energy density} we mean $W = \widetilde{W}(\bE,\textrm{Grad}\bE)$, which is a Helmholtz potential, when integrated over $\Omega_0$.}
\begin{equation}
	\Pi[\bu] = \int_{\Omega_0}  \widetilde{W} (\bE, \mathrm{Grad}\bE)  ~\mathrm{d}V -  \int_{ \Gamma_{0}^T} \bu\cdot \bT \, ~\mathrm{d}S  - \int_{\Gamma_0^M} D \bu \cdot\bM\, ~\mathrm{d}S  - \int_{\Upsilon_0^L} \bu \cdot\bL \, \mathrm{d}C.
\label{eq:totalfreeenergy}
\end{equation}
We recall that the dependence on $\bE$ and $\textrm{Grad}\bE$ renders $\widetilde{W}$ a frame invariant elastic free energy density function for materials of grade two. Here, $\bT$ is the surface traction, $\bM$ is the surface moment and $\bL$ is a line force. Following Equation \eqref{surfacenormalgradient}, $D \bu=(\partial \bu/\partial \bX)\cdot\bN$ is the normal derivative of the displacement on the boundary. Furthermore, $\Gamma_0= \Gamma_{0^i}^u \cup  \Gamma_{0^i}^T = \Gamma_{0^i}^m \cup  \Gamma_{0^i}^M$ represents the decomposition of the smooth surfaces of the boundary and $\Upsilon_0= \Upsilon_{0^i}^l \cup ~\Upsilon_{0^i}^L$ represents the decomposition of the smooth edges of the boundary into Dirichlet subsets (identified by superscripts $u, m ~\textrm{and} ~l$) and Neumann subsets (identified by superscripts $T, M ~\textrm{and} ~L$). We are interested in a displacement field of the following form:
\begin{equation}
u_i \in \mathscr{S}, \;\textrm{such that}\; u_i = \bar{u}_i,\;\forall \bX \in \Gamma_{0^i}^u;\quad u_i = \bar{l}_i,\;\forall \bX\in\Upsilon_{0^i}^l;\quad Du_i = \bar{m}_i,\;\forall \bX\in\Gamma_{0^i}^m
\label{dirbcsu}
\end{equation}

At equilibrium, the first variation of the Gibbs potential with respect to the displacement field is zero. As is standard, to construct such a variation we first consider variations on the displacement field $\bu_\varepsilon := \bu + \varepsilon\bw$, where 
\begin{equation}
 w_i\in\mathscr{V}\;\textrm{such that}\;w_i = 0 ~\forall ~\bX \in \Gamma_{0^i}^u \cup  \Upsilon_{0^i}^l, ~Dw_i = 0 ~\forall ~\bX \in \Gamma_{0^i}^m 
\label{dirbcsw}
\end{equation}
 We construct the first variation of the Gibbs potential with respect to the displacement, adopting coordinate notation for the sake of clarity: 
\begin{align}
  \frac{\delta}{\delta\bu}\Pi[\bu] = &~\frac{\mathrm{d}}{\mathrm{d} \varepsilon} \Pi[\bu_\varepsilon]
   \bigg|_{\varepsilon=0} \nonumber \\
  =  &\int_{\Omega_0} \left( \frac{\partial \widetilde{W}}{\partial E_{AB}}  \frac{\partial E_{AB}}{\partial F_{iJ}} w_{i,J} +  \frac{\partial \widetilde{W}}{\partial E_{AB,C}}\frac{\partial E_{AB,C}}{\partial F_{iJ}}w_{i,J} +  \frac{\partial \widetilde{W}}{\partial E_{AB,C}}  \frac{\partial E_{AB,C}}{\partial F_{iJ,K}}  w_{i,JK} \right) ~\mathrm{d}V  \nonumber \\ 
  -  &\int_{\Gamma_{0^i}^T}  w_{i} T_{i} \, ~\mathrm{d}S  - \int_{\Gamma_{0^i}^M} Dw_i M_{i} \, ~\mathrm{d}S  - \int_{\Upsilon_{0^i}^L} w_{i} L_{i} \, \mathrm{d}C
\end{align}
Here we denote, 
\begin{align}
\frac{\partial \widetilde{W}}{\partial E_{AB}}  \frac{\partial E_{AB}}{\partial F_{iJ}} + \frac{\partial \widetilde{W}}{\partial E_{AB,C}}\frac{\partial E_{AB,C}}{\partial F_{iJ}} = \frac{\partial \widetilde{W}}{\partial F_{iJ}} &= P_{iJ}\label{eqn:stressP} \\
\frac{\partial \widetilde{W}}{\partial E_{AB,C}}  \frac{\partial E_{AB,C}}{\partial F_{iJ,K}} = \frac{\partial \widetilde{W}}{\partial F_{iJ,K}} &=B_{iJK} \label{eqn:stressB}
\end{align}
where $P_{iJ}$ are the components of the first Piola-Kirchhoff stress tensor, and $B_{iJK}$ are the components of the higher-order stress tensor that is conjugate to the higher-order deformation gradient, $F_{iJ,K}$. Note that the symmetry condition $B_{iJK} = B_{iKJ}$ holds. The extremal condition is obtained by setting the first variation of the Gibbs potential to zero, and yields the Euler-Lagrange equation for a material of grade two. We note that with the specification of Equations (\ref{dirbcsu}) and (\ref{dirbcsw}) this also is the weak form of mechanical equilibrium:
\begin{equation}
  \int_{\Omega_0} \left( P_{iJ} w_{i,J} +  B_{iJK} w_{i,JK} \right) ~\mathrm{d}V - \int_{\Gamma_{0^i}^T} w_i T_i \, ~\mathrm{d}S  - \int_{\Gamma_{0^i}^M} Dw_i M_i \, ~\mathrm{d}S  - \int_{\Upsilon_{0^i}^L} w_i L_i \, \mathrm{d}C = 0
\label{eqn:weakform}
\end{equation}

\noindent The fourth-order nature of the problem resides in products of $B_{iJK}$ and $w_{i,JK}$, each of which involves second-order spatial gradients. Before proceeding to the numerical formulation (Section \ref{sec:numerical}), we derive the strong form of the problem. Applying integration by parts to Equation \eqref{eqn:weakform} we obtain, 
\begin{align}
  &- \int_{\Omega_0} P_{iJ,J} w_{i} ~\mathrm{d}V - \int_{\Omega_0} B_{iJK,K} w_{i,J} ~\mathrm{d}V +  \int_{\Gamma_0} P_{iJ}w_{i}N_{J} ~\mathrm{d}S +  \int_{\Gamma_0} B_{iJK}w_{i,J}N_{K} ~\mathrm{d}S \nonumber \\
 &- \int_{\Gamma_{0^i}^T} w_i T_i \, ~\mathrm{d}S  - \int_{\Gamma_{0^i}^M} Dw_i M_i \, ~\mathrm{d}S  - \int_{\Upsilon_{0^i}^L} w_i L_i \, \mathrm{d}C = 0
\label{eqn:weakform2}
\end{align}
Applying integration by parts again, but only on the second volume integral, yields, 
\begin{align}
  &- \int_{\Omega_0} P_{iJ,J} w_{i} ~\mathrm{d}V + \int_{\Omega_0} B_{iJK,JK} w_{i} ~\mathrm{d}V - \underbrace{\int_{\Gamma_0} B_{iJK,K}w_{i}N_{J} ~\mathrm{d}S}_{\text{Integral A}} +  \int_{\Gamma_0} P_{iJ}w_{i}N_{J} ~\mathrm{d}S  \nonumber \\
 & +  \underbrace{\int_{\Gamma_0} B_{iJK}w_{i,J}N_{K} ~\mathrm{d}S}_{\text{Integral B}} - \int_{\Gamma_{0^i}^T} w_i T_i \, ~\mathrm{d}S  - \int_{\Gamma_{0^i}^M} Dw_i M_i \, ~\mathrm{d}S  - \int_{\Upsilon_{0^i}^L} w_i L_i \, \mathrm{d}C = 0
\label{eqn:weakform3}
\end{align}
Expanding the term labelled as Integral A by repeated use of Equation \eqref{surfacenormalgradient},
\begin{align}
 \int_{\Gamma_0} B_{iJK,K}w_{i}N_{J} ~\mathrm{d}S &=  \int_{\Gamma_0} \left( B_{iJK,L} \delta_{LK} \right) w_{i}N_{J} ~\mathrm{d}S \nonumber \\
 &=  \int_{\Gamma_0} \left( D B_{iJK} N_{L} + D_{L} B_{iJK} \right) \delta_{LK} N_{J} w_{i} ~\mathrm{d}S \nonumber \\
 &=  \int_{\Gamma_0} \left( D B_{iJK} N_{K} N_{J} +  D_{K} B_{iJK} N_{J} \right) w_{i} ~\mathrm{d}S.
\label{eqn:weakformA}
\end{align}

\noindent Likewise expanding the term labelled as Integral B
\begin{align}
\int_{\Gamma_0} B_{iJK}w_{i,J}N_{K} ~\mathrm{d}S &=  \int_{\Gamma_0} \left( D w_{i} N_{J} + D_{J} w_{i} \right) B_{iJK} N_{K} ~\mathrm{d}S \nonumber \\
 &=  \int_{\Gamma_0} D w_{i} B_{iJK} N_{J} N_{K} ~\mathrm{d}S  +   \underbrace{\int_{\Gamma_0} D_{J} w_{i} B_{iJK} N_{K} ~\mathrm{d}S}_{\text{Integral C}}.
\label{eqn:weakformB}
\end{align}

\noindent Integral C can be expanded as
\begin{align}
\int_{\Gamma_0} D_{J} w_{i} B_{iJK} N_{K} ~\mathrm{d}S &= \int_{\Gamma_0} D_{J} \left( w_{i} B_{iJK} N_{K} \right) ~\mathrm{d}S  -  \int_{\Gamma_0} w_{i} D_{J} \left( B_{iJK} N_{K} \right) ~\mathrm{d}S \nonumber \\
&= \underbrace{\int_{\Gamma_0} D_{J} \left( w_{i} B_{iJK} N_{K} \right) ~\mathrm{d}S}_{\text{Integral D}}  -  \int_{\Gamma_0} w_{i} \left( D_{J} \left( B_{iJK} \right) N_{K} +  B_{iJK}D_{J}N_{K} \right) ~\mathrm{d}S
\label{eqn:weakformC}
\end{align}
Using the integral identity $\int_{\Gamma_0} D_{I} f_{....} N_{J} ~\mathrm{d}S = \int_{\Gamma_0} (b^K_K N_I N_J - b_{IJ}) f_{....} ~\mathrm{d}S + \int_{\Upsilon_0}  \llbracket N^{\Gamma}_{I} N_{J} f_{....} \rrbracket  ~\mathrm{d}L$ \citep{Toupin1962}, where $b_{IJ}=-D_{I}N_J=-D_{J}N_I$ are components of the second fundamental form of the smooth parts of the boundary and $\bN^\Gamma = \BXi\times\bN$, where $\BXi$ is the unit tangent to the curve $\Upsilon_0$,  Integral D yields
\begin{equation}
\int_{\Gamma_0} D_{J} \left( w_{i} B_{iJK} N_{K} \right) ~\mathrm{d}S = \int_{\Gamma_0}  (b^L_L N_J N_K - b_{JK}) w_{i} B_{iJK} ~\mathrm{d}S + \int_{\Upsilon_0} \llbracket N^{\Gamma}_{J} N_{K} w_{i} B_{iJK} \rrbracket  ~\mathrm{d}L.
\label{eqn:weakformD}
\end{equation}
Collecting terms from Equations (\ref{eqn:weakform3}--\ref{eqn:weakformD}), and using $B_{iJK} = B_{iKJ}$,
\begin{align}
  -& \int_{\Omega_0} w_{i} \left(P_{iJ,J} - B_{iJK,JK} \right)~\mathrm{d}V \nonumber \\
  &+ \int_{\Gamma_0} w_{i} \left(P_{iJ}N_J - DB_{iJK}N_KN_J - 2D_J(B_{iJK})N_K - B_{iJK}D_JN_K + (b^L_LN_JN_K-b_{JK})B_{iJK} \right)~\mathrm{d}S \nonumber \\
  &+ \int_{\Gamma_0} Dw_iB_{iJK}N_JN_K ~\mathrm{d}S  \nonumber \\
  &+ \int_{\Upsilon_0} w_i \llbracket N^{\Gamma}_{J} N_{K} B_{iJK} \rrbracket  ~\mathrm{d}L \nonumber \\
  &- \int_{\Gamma_{0^i}^T} w_i T_i \, ~\mathrm{d}S  - \int_{\Gamma_{0^i}^M} Dw_i M_i \, ~\mathrm{d}S  - \int_{\Upsilon_{0^i}^L} w_i L_i \, \mathrm{d}C = 0
\label{eqn:weakformFinal}
\end{align}

Standard variational arguments, including the invocation of homogeneous boundary conditions on $w_i$ on $\Gamma_{0_i}^u\cup\Upsilon_{0_i}^l$ and $Dw_i$ on $\Gamma_{0_i}^m$, then lead to the strong form of mechanical equilibrium for a material of grade two: \\
\begin{equation}
\begin{array}{lll}
P_{iJ,J} - B_{iJK,JK} &= 0 &\mathrm{in} ~\Omega_0\\
u_{i}  &= \bar{u}_i   &\mathrm{on} ~\Gamma_{0^i}^u\\
P_{iJ}N_J - DB_{iJK}N_KN_J - 2D_J(B_{iJK})N_K - B_{iJK}D_JN_K + (b^L_LN_JN_K-b_{JK})B_{iJK} & = T_{i} &\mathrm{on} ~\Gamma_{0^i}^T\\
Du_i  &= \bar{m}_i &\mathrm{on} ~\Gamma_{0^i}^m\\
B_{iJK}N_JN_K &= M_{i} &\mathrm{on}  ~\Gamma_{0^i}^M\\
u_{i}  &= \bar{l}_i  &\mathrm{on} ~\Upsilon_{0^i}^l\\
\llbracket N^{\Gamma}_{J} N_{K} B_{iJK} \rrbracket &= L_{i} &\text{\small on} ~\Upsilon_{0^i}^L\\ \\
 \textrm{where,}\qquad \Gamma_0= \Gamma_{0^i}^u \cup  \Gamma_{0^i}^T, \qquad \Gamma_0= \Gamma_{0^i}^m \cup  \Gamma_{0^i}^M, \qquad \Upsilon_0= \Upsilon_{0^i}^l \cup  \Upsilon_{0^i}^L&
\end{array}
\label{eqn:strongformgradelasticity}
\end{equation}

The (nonlinear) fourth-order nature of the governing partial differential equation above is clarified by noting that $B_{iJK,JK}$ introduces $F_{aB,CJK}$ via Equation \eqref{eqn:stressB}. The Dirichlet boundary condition in \eqref{eqn:strongformgradelasticity}$_2$ has the same form as for conventional elasticity. However, its dual Neumann boundary condition, \eqref{eqn:strongformgradelasticity}$_3$ is notably more complex than its conventional counterpart, which would have only the first term on the left hand-side. Equation \eqref{eqn:strongformgradelasticity}$_4$ is the higher-order Dirichlet boundary condition applied to the normal gradient of the displacement field, and Equation \eqref{eqn:strongformgradelasticity}$_5$ is the higher-order Neumann boundary condition on the higher-order stress, $\bB$. Adopting the physical interpretation of $\bB$ as a couple stress \citep{Toupin1962}, the homogeneous form of this boundary condition, if extended to the atomic scale, states that there is no boundary mechanism to impose a generalized moment across atomic bonds. Finally, Equation \eqref{eqn:strongformgradelasticity}$_6$ is the Dirichlet boundary condition on the smooth edges of the boundary and Equation \eqref{eqn:strongformgradelasticity}$_7$ is its conjugate Neumann boundary condition. Following \cite{Toupin1962}, the homogeneous form of this condition requires that there be no discontinuity in the higher order (couple) stress traction across a smooth edge $\Upsilon_0^L$ in the absence of a balancing line traction along $\Upsilon_0^L$.

%
%
\section{Numerical treatment}
\label{sec:numerical}
\subsection{Weak form of the continuous problem}
For completeness we restate the weak form: Find $u_i \in \mathscr{S}$, where $\mathscr{S}= \{ u_i ~\vert  ~u_{i} = ~\bar{u}_i ~\forall ~\bX \in \Gamma_{0^i}^u, ~Du_i = ~\bar{m}_i ~\forall ~\bX \in \Gamma_{0^i}^m, ~u_{i} = ~\bar{l}_i  ~\forall ~\bX \in \Upsilon_{0^i}^l \}$,  such that $\forall ~w_i \in \mathscr{V}$, where $\mathscr{V}= \{ w_i ~\vert  ~w_{i} = ~0 ~\forall ~\bX \in \Gamma_{0^i}^u, ~Dw_i = 0 ~\forall ~\bX \in \Gamma_{0^i}^m, ~w_{i} = ~0  ~\forall ~\bX \in \Upsilon_{0^i}^l \}$
\begin{equation}
  \int_{\Omega_0} \left( P_{iJ} w_{i,J} +  B_{iJK} w_{i,JK} \right) ~\mathrm{d}V - \int_{\Gamma_{0^i}^T} w_i T_i \, ~\mathrm{d}S  - \int_{\Gamma_{0^i}^M} Dw_i M_i \, ~\mathrm{d}S  - \int_{\Upsilon_{0^i}^L} w_i L_i \, \mathrm{d}C = 0
\label{eqn:continuousproblem}
\end{equation}

\subsection{Galerkin formulation}
\label{sec:galerkin}

As always, the Galerkin weak form is obtained by restriction to finite dimensional functions $(\bullet)^h$: Find $u^h_i \in \mathscr{S}^h \subset \mathscr{S}$, where $\mathscr{S}^h= \{ u^h_i \in \mathscr{H}^2(\Omega_0) ~\vert  ~u^h_{i} = ~\bar{u}_i ~\forall ~\bX \in \Gamma_{0^i}^u, ~Du^h_i = ~\bar{m}_i ~\forall ~\bX \in \Gamma_{0^i}^m, ~u^h_{i} = ~\bar{l}_i  ~\forall ~\bX \in \Upsilon_{0^i}^l \}$,  such that $\forall ~w^h_i \in \mathscr{V}^h \subset \mathscr{V}$, where $\mathscr{V}^h= \{ w^h_i \in\mathscr{H}^2(\Omega_0)~\vert  ~w^h_{i} = ~0 ~\forall ~\bX \in \Gamma_{0^i}^u, ~Dw^h_i = 0 ~\forall ~\bX \in \Gamma_{0^i}^m, ~w^h_{i} = ~0  ~\forall ~\bX \in \Upsilon_{0^i}^l \}$
\begin{equation}
  \int_{\Omega_0} \left( P^h_{iJ} w^h_{i,J} +  B^h_{iJK} w^h_{i,JK} \right) ~\mathrm{d}V - \int_{\Gamma_{0^i}^T} w^h_i T_i \, ~\mathrm{d}S  - \int_{\Gamma_{0^i}^M} Dw^h_i M_i \, ~\mathrm{d}S  - \int_{\Upsilon_{0^i}^L} w^h_i L_i \, \mathrm{d}C = 0
\label{eqn:galerkinform}
\end{equation}

\noindent The second-order gradients in the weak form require the solutions to lie in $\mathscr{H}^2(\Omega_0)$, a more restrictive condition than the formulation of finite strain elasticity for materials of grade one where the solutions are drawn from the larger space $\mathscr{H}^1(\Omega_0) \supset \mathscr{H}^2(\Omega_0)$.The variations, $\bw^h$ and trial solutions $\bu^h$ are defined component-wise using a finite number of basis functions,
\begin{equation}
w_i^h = \sum_{a=1}^{n_\mathrm{b}} w_i^a N^a, \quad \qquad u_i^h = \sum_{a=1}^{n_\mathrm{b}} u_i^a N^a 
\label{eq:basisdef}
\end{equation}
\noindent where $n_\mathrm{b}$ is the dimensionality of the function spaces $\mathscr{S}^h$ and $\mathscr{V}^h$, and $N^a$ represents the basis functions. Since $\mathscr{S}^h \subset \mathscr{H}^2$, $C^0$ basis functions do not provide the required degree of regularity demanded by the problem; however, it suffices to consider $C^1$ basis functions in $\mathscr{S}^h$. One possibility is the use of $C^1$ Hermite elements as in \cite{Papanicolopulos2009}. Alternately, one could invoke the class of continuous/discontinuous Galerkin methods \cite{Engeletal2002, Wellsetal2004, Molarietal2006}, in which the displacement field is $C^0$-continuous, but the strains are discontinuous  across element interfaces. This class of methods is more complex, and has additional stability requirements. A mixed formulation of finite strain gradient elasticity could be constructed by introducing an independent kinematic field for the deformation gradient or another strain measure. However, this approach leads to boundary conditions that do not admit straightforward interpretations.  We prefer to avoid the complexities of Hermite elements in three dimensions, and seek to circumvent the challenges posed by discontinuous Galerkin methods and mixed formulations by turning to Isogeometric Analysis introduced by \citet{HughesCottrellBazilevs2005}. Also see \citet{CottrellHughesBazilevs2009} for details.\\

\subsubsection{Isogeometric Analysis}
\label{sec:iga}
As is now well-appreciated in the computational mechanics community, Isogeomeric Analysis (IGA) is a mesh-based numerical method with NURBS (Non-Uniform Rational B-Splines) basis functions. The NURBS basis leads to many desirable properties, chief among them being the exact representation of the problem geometry. Like the Lagrange polynomial basis functions traditionally used in the Finite Element Method (FEM), the NURBS basis functions are partitions of unity with compact support, satisfy affine covariance (i.e an affine transformation of the basis is obtained by the affine transformation of its nodes/control points) and support an isoparametric formulation, thereby making them suitable for a Galerkin framework. They enjoy advantages over Lagrange polynomial basis functions in being able to ensure $C^n$-continuity, in possessing the positive basis and convex hull properties, and being variation diminishing. A detailed discussion of the NURBS basis and IGA is beyond the scope of this article and interested readers are referred to \citet{CottrellHughesBazilevs2009}. However, we briefly present the construction of a $C^1$-continuous NURBS basis. \\

\begin{figure}[h]
  \centering
   \psfrag{a}[][][0.75]{$B^1$}
   \psfrag{b}[][][0.75]{$B^2$}
   \psfrag{c}[][][0.75]{$B^3$}
   \psfrag{d}[][][0.75]{$B^4$}
   \psfrag{e}[][][0.75]{$B^5$}
   \psfrag{f}[][][0.75]{$B^6$}
   \psfrag{g}[][][0.75]{$B^7$}
   \psfrag{h}[][][0.75]{$B^8$}
   \psfrag{i}[][][0.75]{$\xi$}
  \includegraphics[width=0.6\textwidth]{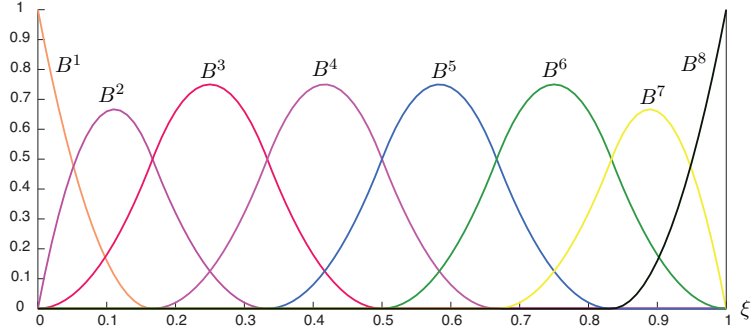}
\caption{Quadratic (p=2) B-spline basis constructed from the knot vector $\chi=$ [0, 0 , 0 , 1/6, 1/3, 1/2, 2/3, 5/6, 1, 1, 1].}
\label{fig:bsplines}
\end{figure}

The building blocks of the NURBS basis functions are univariate B-spline functions that are defined as follows: Consider two positive integers $p$ and $n$, and a non-decreasing sequence of values $\chi=[\xi_1, \xi_2,...., \xi_{n+p+1}]$, where p is the polynomial order, n is the number of basis functions, the $\xi_i$ are coordinates in the parametric space referred to as knots (equivalent to nodes in FEM) and $\chi$ is the knot vector. The B-spline basis functions $B_{i,p}(\xi)$ are defined starting with the zeroth order basis functions
\begin{align}
B_{i,0}(\xi) &= \left\{\begin{array}{ll}
1 &\mathrm{if}\;\xi_i \le \xi < \xi_{i+1},\\
0 &\mathrm{otherwise}
\end{array}\right.
\end{align}
and using the Cox-de Boor recursive formula for $p \geq 1$ \citep{Piegl1997}
\begin{align}
 B_{i,p} (\xi) &=
  \frac{\xi-\xi_i}{\xi_{i+p}-\xi_i} B_{i,p-1} (\xi) + \frac{\xi_{i+p+1}-\xi}{\xi_{i+p+1}-\xi_{i+1}} B_{i+1,p-1} (\xi)
\end{align}
The knot vector divides the parametric space into intervals referred to as knot spans (equivalent to elements in FEM). A B-spline basis function is $C^{\infty}$-continuous inside knot spans and $C^{p-1}$-continuous at the knots. If an interior knot value repeats, it is referred to as a multiple knot. At a knot of multiplicity $k$, the continuity is $C^{p-k}$. Now, using a quadratic B-spline basis (Figure (\ref{fig:bsplines})), a $C^1$-continuous one dimensional NURBS basis is given by
\begin{align}
N^{i} (\xi) =
  \frac{B_{i,2} (\xi) \textit{w}_{i}}{\sum_{i=1}^{n_b} B_{i,2} (\xi) \textit{w}_{i}}
\end{align}
where $w_i$ are the weights associated with each of the B-spline functions. In higher-dimensions, NURBS basis functions are constructed as a tensor product of the one dimensional basis functions:
\begin{align}
 N^{ij} (\xi,\eta) &=
  \frac{B_{i,2} (\xi) B_{j,2} (\eta) \textit{w}_{ij}}{\sum_{i=1}^{n_{b1}} \sum_{j=1}^{n_{b2}} B_{i,2}(\xi) B_{j,2} (\eta) \textit{w}_{ij}} &\mathrm{(2D)} \nonumber \\
 N^{ijk} (\xi,\eta, \zeta) &=
  \frac{B_{i,2} (\xi) B_{j,2} (\eta) B_{k,2} (\zeta) \textit{w}_{ijk}}{\sum_{i=1}^{n_{b1}} \sum_{j=1}^{n_{b2}} \sum_{k=1}^{n_{b3}} B_{i,2}(\xi) B_{j,2} (\eta) B_{k,2} (\zeta) \textit{w}_{ijk}} &\mathrm{(3D)}
\label{eq:higherordernurbs}
\end{align}

\subsubsection{Higher-order Dirichlet boundary conditions}
\label{sec:dirichlet}
The enforcement of the higher-order Dirichlet boundary condition (Equation \eqref{eqn:strongformgradelasticity}$_4$) encountered in a gradient elasticity formulation poses numerical challenges. Usually Dirichlet boundary conditions are specified on the primal fields and are numerically enforced by building the boundary condition into the finite dimensional function space, for example, $\mathscr{S}^h \subset \mathscr{S} = \{ u_i ~\vert ~u_i = ~\bar{u}_i ~\forall ~\bX \in \Gamma^u_{0^i} \} $, where $\Gamma_0$. A similar approach to enforcing the higher-order Dirichlet boundary conditions involving the normal gradient of the primal field would require the construction of a finite dimensional function space, $\mathscr{S}^h \subset \mathscr{S} = \{ u_i ~\vert ~Du_i = ~\bar{m}_i ~\forall ~\bX \in \Gamma^m_{0^i} \}$, which often may not be possible using the standard finite element basis. In specific cases, one could constrain the nodes/control-points in the first layer of elements/knot-spans adjacent to the Dirichlet boundary to enforce the higher-order Dirichlet boundary condition, but this approach presents complications for arbitrary meshes. Hence we propose a modified Galerkin formulation of the continuous problem (Equation \ref{eqn:continuousproblem}) that weakly enforces the higher-order Dirichlet boundary condition using a penalty-based approach: \\ \\
Find $u_i^h \in \mathscr{S}^h$, where $\mathscr{S}^h= \{ u_i^h ~\vert ~u_i^h \in \mathscr{H}^2(\Omega_0), ~u^h_{i} = ~\bar{u}_i ~\forall ~\bX \in \Gamma_{0^i}^u, ~u^h_{i} = ~\bar{l}_i  ~\forall ~\bX \in \Upsilon_{0^i}^l \}$,  such that $\forall ~w_i^h \in \mathscr{V}^h$,
\begin{equation}
\begin{array}{rcl}
&\int_{\Omega_0} \left( P^h_{iJ} w^h_{i,J} +  B^h_{iJK} w^h_{i,JK} \right) ~\mathrm{d}V - \int_{\Gamma_{0^i}^T} w^h_i T_i \, ~\mathrm{d}S  - \int_{\Gamma_{0^i}^M} Dw^h_i M_i \, ~\mathrm{d}S  - \int_{\Upsilon_{0^i}^L} w^h_i L_i \, \mathrm{d}C \\ \\
&- \int_{\Gamma_{0^i}^m} Dw^h_i \left( B^h_{iJK}N_JN_K \right) \, ~\mathrm{d}S + \frac{C}{h_e} \int_{\Gamma_{0^i}^m} Dw^h_i\left(Du^h_i - \bar{m}_i \right) \, ~\mathrm{d}S =0
\end{array}
\label{eqn:weakformgradelasticity}
\end{equation}
The exact solution $u_i$  satisfies Equation (\ref{eqn:weakformgradelasticity}), hence this weak form is consistent. Here the last two terms enforce consistency, and the higher-order Dirichlet boundary condition via a penalty, respectively. The derivation of the strong form corresponding to the original weak formulation [Equation (\ref{eqn:weakform})] yields a term that is canceled by the consistency term. Here, $C$ is a positive penalty parameter and $h_e$ is the characteristic mesh size parameter.\footnote{The optimal value of $C$ is dependent on the polynomial order of interpolation and the element type \citep{Ciarlet1978}. However, to generate results for this paper $C=5$ was chosen.}  This approach to enforcing the Dirichlet boundary condition weakly by adding a consistency term (along with an adjoint consistency term for symmetric problems) and a penalty term was motivated by \citet{Nitsche1971} and is often used in discontinuous Galerkin methods \citep{Arnold2001}. In the context of IGA, it was introduced by \citet{Bazilevs2007} to weakly enforce Dirichlet boundary conditions for boundary layer solutions of the advection-diffusion equation and incompressible Navier-Stokes equation.
 
\subsubsection{Algorithmic differentiation}
\label{sec:auto}
The residual equation for the solution procedure is given by Equation (\ref{eqn:weakformgradelasticity}), which is highly nonlinear in the primal field, $\bu$. The extent of this non-linearity can be appreciated by expanding the expressions for the stress terms [Equations (\ref{eqn:stressP}, \ref{eqn:stressB})] in terms of gradients of $u_i$. As a result, the analytical linearization of this residual equation to obtain the Jacobian matrix is tedious and is fraught with the danger of algebraic mistakes. Symbolic differentiation is an option, but the computational cost involved is prohibitively high. A standard alternative is the use of numerical differentiation tools built into many standard solver packages. However, for a highly non-linear set of equations, numerical differentiation is inaccurate and ultimately unstable. An effective and efficient alternative is the use of algorithmic (or automatic) differentiation (AD). The key insight to AD is that every equation in a computer program, no matter how complex, ultimately involves a sequence of elementary arithmetic operations and elementary function evaluations (polynomial, trigonometric, logarithmic, exponential or reciprocal). Thus, by the repeated application of the chain rule, any equation can be differentiated by reduction to arithmetic operations and elementary function evaluations, which can be performed with machine precision. The computational cost is at most a small constant times the cost of evaluation of the original equations. We use AD in this work to linearize Equation (\ref{eqn:weakformgradelasticity}) and compute the Jacobian matrix. Specifically, we use the Sacado package, which is part of the open-source Trilinos project ~\citep{Sacado2005,sacado2012}.

%
%
\section{Numerical Simulations}
\label{sec:simulations}

We restrict ourselves to numerical examples of classical boundary value problems in this communication. The familiar setting of these problems makes it somewhat easier to appreciate the role of strain gradient elasticity. However, in Section \ref{sec:discussion}, we point to the truly compelling applications for our framework with regard to martensitic phase transformations, crack tip fields and dislocation cores.

We consider the following material model 
\begin{equation}
\widetilde{W} (\bE, \mathrm{Grad}\bE)  = \frac{\lambda}{2} \left( E_{AA} \right)^2 + \mu  \left( E_{AB} E_{AB} \right) + \frac{1}{2} \mu l^2 E_{AB,C} E_{AB,C} 
\label{eq:quadraticelasticenergy}
\end{equation}
where the first two terms represent the strain-dependent component of the elastic energy density given by the standard St. Venant-Kirchhoff model and the last term is the strain gradient-dependent component, for which we have chosen a quadratic form. The Lam{\'e} parameters are $\lambda$ and $\mu$, and $l$ is the gradient length scale parameter. For the infinitesimal strain theory, the strain gradient component of the above elastic energy density reduces to the quadratic form $\varepsilon_{ab,c}\varepsilon_{ab,c}$ of the material model proposed by \citet{Mindlin1964}. Substituting Equation (\ref{eq:quadraticelasticenergy}) in Equations (\ref{eqn:stressP}, \ref{eqn:stressB}) we obtain the stress measures as
\begin{align}
 P_{iJ} &= \lambda E_{AA} F_{iJ} + 2 \mu F_{iA}E_{AJ} + \mu l^2 E_{AJ,C} F_{iA,C} \label{eqn:stressP1} \\
 B_{iJK} &= \mu \l^2 F_{iA} E_{AJ,K} \label{eqn:stressB1}
\end{align}
Numerical solutions follow to a wide range of boundary value problems across spatial dimensions, starting with the validation of the numerical treatment by comparing the analytical and numerical solutions for uniaxial tension. 

\subsection{Comparison of analytical and numerical solutions for uniaxial tension in one dimension}
\label{sec:1D}
We consider the following problem \citep{Engeletal2002}, which is the one dimensional reduction of the strong form in Equation \eqref{eqn:strongformgradelasticity} to infinitesimal strain with the material model of Equation (\ref{eq:quadraticelasticenergy}) for $\lambda=0$:
\begin{equation}
\begin{array}{rcl}
\sigma_{,x} - \beta_{,xx} &= 0 \quad &\forall \quad x \in (0, L) \\
u  &= 0   \quad &\mathrm{at} \quad  x={0} \\
\left(\sigma - \beta \right).n  &= t  \quad &\mathrm{at} \quad x={L}\\
u_{,x}.n &= 0 \quad &\mathrm{at} \quad x=\{0,L\}
\end{array}
\label{eqn:strongform1D}
\end{equation}
where $\sigma=\mu u_{,x}$, $\beta=\mu l^2 u_{,xx}$, $n = -1$ at $x = 0$ and $n = 1$ at $x = L$. The exact solution to this problem is given by
\begin{equation}
u(x) = \frac{t~l}{\mu \left( e^{L/l} + 1\right)} \left( 1 - e^{L/l} +  e^{(L-x)/l} - e^{x/l} \right) + \frac{t}{\mu}x
\label{eqn:analyticalSolution1D}
\end{equation}
To validate the Galerkin formulation, its discretization using the $C^n$ B-spline basis functions, and linearization using algorithmic differentiation, we compare the above analytic solution to the corresponding numerical solutions obtained under the assumption of infinitesimal strain. First, we demonstrate the effect of the degree of continuity of basis functions on the numerical solution in Figure \ref{fig:OneDAnalytical}. It is easy to check that $C^0$ basis functions lack the degree of high-order continuity demanded by the exact solution, and the corresponding numerical solution significantly differs from the analytic solution when gradient effects become significant. However, $C^n$ basis functions guarantee the required continuity of solution for $n \ge 1$ and thus accurately resolve the analytic solution for all values of the gradient length scale parameter. We draw attention to the fact that, at low values of the gradient length scale parameter, $l$, the strains are large (note the gradient of the displacement field), while the strain gradients are negligible (second gradient of the displacement field). However, as $l$ increases, the strain gradients become more pronounced near the boundaries, $x = \{0,L\}$, and make the overall response stiffer, if viewed in terms of the maximum displacement. Furthermore, optimal convergence rates of $p+1-m$ are attained in the $\mathscr{H}^m$ semi-norm, where $p$ is the polynomial order of basis functions. This has been demonstrated with respect to the solution's $\mathscr{H}^1$ and $\mathscr{H}^2$ semi-norms as shown in Figure \ref{fig:OneDNorms}. 

As noted in the Introduction, the analytical and numerical complexities that arise in the study of strain gradient elasticity have restricted the advances to infinitesimal strain formulations. To emphasize the importance of finite strain formulations of gradient elasticity, we present a comparison of the numerical solutions (displacement, strain and strain gradient) to the infinitesimal strain and finite strain formulations of this problem in Figure \ref{fig:OneDFiniteStrain}. We note that the interaction of the nonlinearity and strain gradients renders a much stiffer result with the finite strain formulation. For this reason, problems involving strong strain gradient effects may suffer a significant loss of accuracy when modeled with infinitesimal strains than with the full finite strain formulation. 

These results validate the numerical treatment presented here. Using this treatment we have obtained solutions to various boundary value problems in higher spatial dimensions, which have been presented in the following sections.

\pgfplotstableread{linearQuadraticCubicl01l1l10Forh0.01FSfalse.u.dat}{\uFSfalseNURBSValues}
\pgfplotstableread{linearQuadraticCubicl01l1l10Forh0.01FStrue.u.dat}{\uFStrueNURBSValues}
\pgfplotstableread{linearQuadraticCubicl01l1l10Forh0.01FSfalse.strain.dat}{\strainFSfalseNURBSValues}
\pgfplotstableread{linearQuadraticCubicl01l1l10Forh0.01FStrue.strain.dat}{\strainFStrueNURBSValues}
\pgfplotstableread{linearQuadraticCubicl01l1l10Forh0.01FSfalse.gradient.dat}{\gradientFSfalseNURBSValues}
\pgfplotstableread{linearQuadraticCubicl01l1l10Forh0.01FStrue.gradient.dat}{\gradientFStrueNURBSValues}
\pgfplotstableread{analyticalSoln.dat}{\analyticalValues}
\pgfplotstableread{tension3DValues.dat}{\tensionValues}
\pgfplotstableread{bending3DValues.dat}{\bendingValues}
\pgfplotstableread{torsion3DValues.dat}{\torsionValues}
\pgfplotstableread{torsionTheta3DValuesl10Weak1FStrue.dat}{\torsionThetaValues}
\pgfplotstableread{oneDNorms_lam-0.00mu-0.50muSG-1.00l-1.00load-1.00G-0.00C-5.00Weak-1FS-false.dat}{\oneDNorms}
\pgfplotstableread{lineLoad.dat}{\lineLoad}
\begin{figure}[h]
  \centering
  \subfloat[$l=0.01$\label{fig:OneDAnalyticall001}]{
    \begin{tikzpicture}[scale=0.55]
      \begin{axis}[minor tick num=1,xlabel={\Large $x$},ylabel={\Large $u$}, x unit=m, y unit=m, mark repeat={10}] 
        \addplot [red, ultra thick] table [x={x}, y={l001_u}] {\analyticalValues};
        \addplot [blue, mark=o, mark size=3] table [x={xval}, y={lel100_l001}] {\uFSfalseNURBSValues};
        \addplot [orange, mark=square, mark size=3] table [x={xval}, y={qel100_l001}] {\uFSfalseNURBSValues};
        \addplot [teal, mark=x, mark size= 5] table [x={xval}, y={cel100_l001}] {\uFSfalseNURBSValues};
      \end{axis}
    \end{tikzpicture}
  }
  \subfloat[$l=0.1$\label{fig:OneDAnalyticall01}]{
    \begin{tikzpicture}[scale=0.55]
      \begin{axis}[minor tick num=1,xlabel={\Large $x$},ylabel={\Large $u$}, x unit=m, y unit=m, mark repeat={10}] 
        \addplot [red, ultra thick] table [x={x}, y={l01_u}] {\analyticalValues};
        \addplot [blue, mark=o,  mark size=3] table [x={xval}, y={lel100_l01}] {\uFSfalseNURBSValues};
        \addplot [orange, mark=square, mark size=3] table [x={xval}, y={qel100_l01}] {\uFSfalseNURBSValues};
        \addplot [teal, mark=x, mark size= 5] table [x={xval}, y={cel100_l01}] {\uFSfalseNURBSValues};
      \end{axis}
    \end{tikzpicture}
  }
  \subfloat[$l=1.0$\label{fig:OneDAnalyticall1}]{
    \begin{tikzpicture}[scale=0.55]
      \begin{axis}[minor tick num=1,xlabel={\Large $x$},ylabel={\Large $u$}, x unit=m, y unit=m, mark repeat={10}] 
        \addplot [red, ultra thick] table [x={x}, y={l1_u}] {\analyticalValues};
        \addplot [blue,  mark=o,  mark size=3] table [x={xval}, y={lel100_l1}] {\uFSfalseNURBSValues};
        \addplot [orange, mark=square, mark size= 3] table [x={xval}, y={qel100_l1}] {\uFSfalseNURBSValues};
        \addplot [teal, mark=x, mark size= 5] table [x={xval}, y={cel100_l1}] {\uFSfalseNURBSValues};
        \legend{{\Large Analytical}, {\Large $C^0$ basis}, {\Large $C^1$ basis}, {\Large $C^2$ basis}}
      \end{axis}
    \end{tikzpicture}
  }
  \caption{Comparison of the exact solution (Equation \eqref{eqn:analyticalSolution1D} with $\mu=1.0$, $t=1.0$, $L=1.0$) and corresponding numerical solutions with $C^0$, $C^1$ and $C^2$ bases for different values of the gradient length scale parameter $l$. In each case, the discretization consisted of 100 uniform knot spans.}
  \label{fig:OneDAnalytical}
\end{figure}
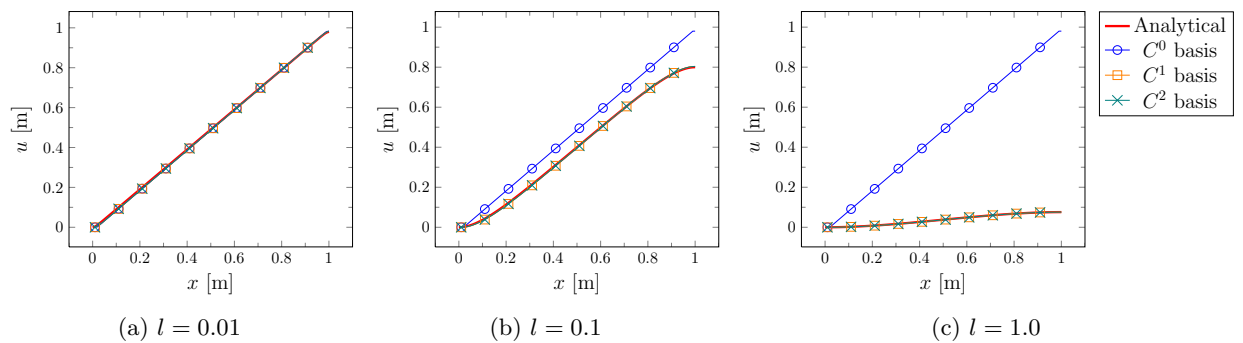

\begin{figure}[h]
  \centering
  \subfloat[$\mathscr{H}^1$ semi-norm\label{fig:h1Norm}]{
    \begin{tikzpicture}[scale=0.65]
      \begin{loglogaxis}[minor tick num=1,xlabel={\Large $h$},ylabel={\Large $\vert u-u^h\vert_{\mathscr{H}^1}$}] 
        \addplot [red, mark=square, thick] table [x={xval}, y={qH1}] {\oneDNorms};
        \addplot [blue, mark=x, thick, mark size=4] table [x={xval}, y={cH1}] {\oneDNorms};
        \addplot [black, no markers] coordinates {(0.01,4.0e-06) (0.01,4.0e-08)};
        \addplot [black, no markers] coordinates {(0.001,4.0e-08) (0.01,4.0e-08)};
        \node[below] at (axis cs:0.003,4.0e-08){\large $1$};
        \node[right] at (axis cs:0.01,4.0e-07){\large $2$};
        \addplot [black, no markers] coordinates {(0.1,5.0e-06) (0.1,4.0e-08)};
        \addplot [black, no markers] coordinates {(0.018,4.0e-08) (0.1,4.0e-08)};
        \node[below] at (axis cs:0.04,4.1e-08){\large $1$};
        \node[right] at (axis cs:0.1,4.1e-07){\large $3$};
      \end{loglogaxis}
    \end{tikzpicture}
  }
  \subfloat[$\mathscr{H}^2$ semi-norm\label{fig:h2Norm}]{
    \begin{tikzpicture}[scale=0.65]
      \begin{loglogaxis}[minor tick num=1,xlabel={\Large $h$},ylabel={\Large$\vert u-u^h\vert_{\mathscr{H}^2}$}] 
        \addplot [red, mark=square, thick] table [x={xval}, y={qH2}] {\oneDNorms};
        \addplot [blue, mark=x, thick, mark size=4] table [x={xval}, y={cH2}] {\oneDNorms};
        \addplot [black, no markers] coordinates {(0.1,2.6693e-02) (0.1,2.6698e-03)};
        \addplot [black, no markers] coordinates {(0.01,2.6698e-03) (0.1,2.6698e-03)};
        \node[below] at (axis cs:0.05,2.6693e-03){\large $1$};
        \node[right] at (axis cs:0.1,1.0e-02){\large $1$};
        \addplot [black, no markers] coordinates {(0.1,1e-04) (0.1,2e-06)};
        \addplot [black, no markers] coordinates {(0.015,2e-06) (0.1,2e-06)};
        \node[below] at (axis cs:0.05,2.0e-06){\large $1$};
        \node[right] at (axis cs:0.1,2.0e-05){\large $2$};
        \legend{{\Large $C^1$ quadratic basis}, {\Large $C^2$ cubic basis}}
      \end{loglogaxis}
    \end{tikzpicture}
  }
  \caption{Optimal convergence of the numerical solutions of the boundary value problem given by Equation \eqref{eqn:strongform1D} ($\mu=1.0$, $l=1.0$, $t=1.0$, $L=1.0$) with respect to mesh discretization ($h=L/\text{number of knot spans}$) in the $\mathscr{H}^1$ and $\mathscr{H}^2$ semi-norms.}
  \label{fig:OneDNorms}
\end{figure}
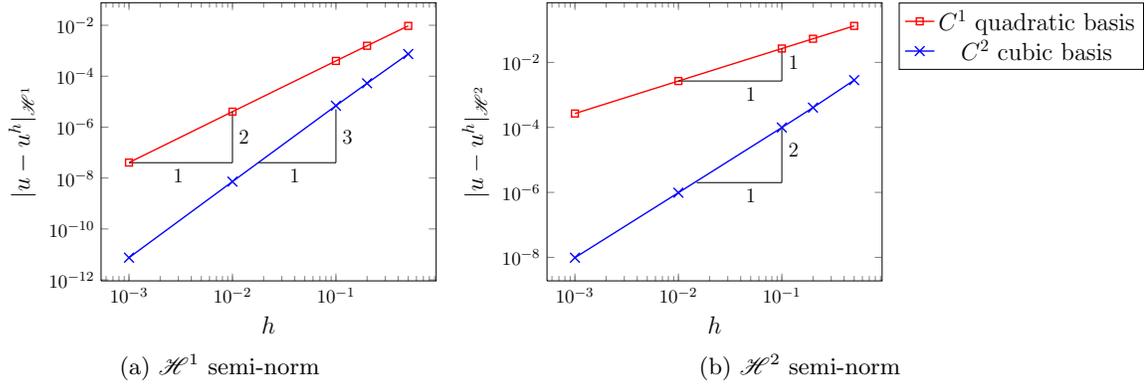

\begin{figure}[h]
  \centering
  \subfloat[$u$\label{fig:OneDFiniteStrainu}]{ 
    \begin{tikzpicture}[scale=0.55]
      \begin{axis}[minor tick num=1,xlabel={\Large $x$},ylabel={\Large $u$}, x unit=m, y unit=m, mark repeat={10}]
        \addplot [red, mark=o, thick, mark size=3] table [x={x}, y={l01_u}] {\analyticalValues};
        \addplot [blue, mark=x, thick, mark size=4] table [x={xval}, y={qel100_l01}] {\uFSfalseNURBSValues};
        \addplot [orange, mark=square, thick, mark size=3] table [x={xval}, y={qel100_l01}] {\uFStrueNURBSValues};
      \end{axis}
    \end{tikzpicture}
  }
  \subfloat[$u_{,x}$\label{fig:OneDFiniteStrainux}]{ 
    \begin{tikzpicture}[scale=0.55]
      \begin{axis}[minor tick num=1,xlabel={\Large $x$},ylabel={\Large $u_{,x}$}, x unit=m, mark repeat={10}]
        \addplot [red, mark=o, thick, mark size=3] table [x={x}, y={l01_ux}] {\analyticalValues};
        \addplot [blue, mark=x, thick, mark size=4] table [x={xval}, y={qel100_l01}] {\strainFSfalseNURBSValues};
        \addplot [orange, mark=square, thick, mark size=3] table [x={xval}, y={qel100_l01}] {\strainFStrueNURBSValues};
      \end{axis}
    \end{tikzpicture}
  }
  \subfloat[$u_{,xx}$\label{fig:OneDFiniteStrainluxx}]{ 
    \begin{tikzpicture}[scale=0.55]
      \begin{axis}[minor tick num=1,xlabel={\Large $x$},ylabel={\Large $u_{,xx}[\text{m}^{-1}]$}, x unit=m, mark repeat={10}, mark size=2pt]
        \addplot [red, mark=o, thick, mark size=3] table [x={x}, y={l01_uxx}] {\analyticalValues};
        \addplot [blue, mark=x,  thick, mark size=4] table [x={xval}, y={qel100_l01}] {\gradientFSfalseNURBSValues};
        \addplot [orange, mark=square, thick, mark size=3] table [x={xval}, y={qel100_l01}] {\gradientFStrueNURBSValues};
        \legend{Analytical  (small strain), $C^1$ basis (small strain), $C^1$ basis (finite strain)}
      \end{axis}
    \end{tikzpicture}
  }
  \caption{Comparison of numerical solutions obtained using the infinitesimal strain assumption (Equation \eqref{eqn:strongform1D} with $\mu=1.0$, $l=0.1$, $t=1.0$, $L=1.0$) and its corresponding finite strain formulation. The discretization consisted of 100 uniform knot spans.}
  \label{fig:OneDFiniteStrain}
\end{figure}
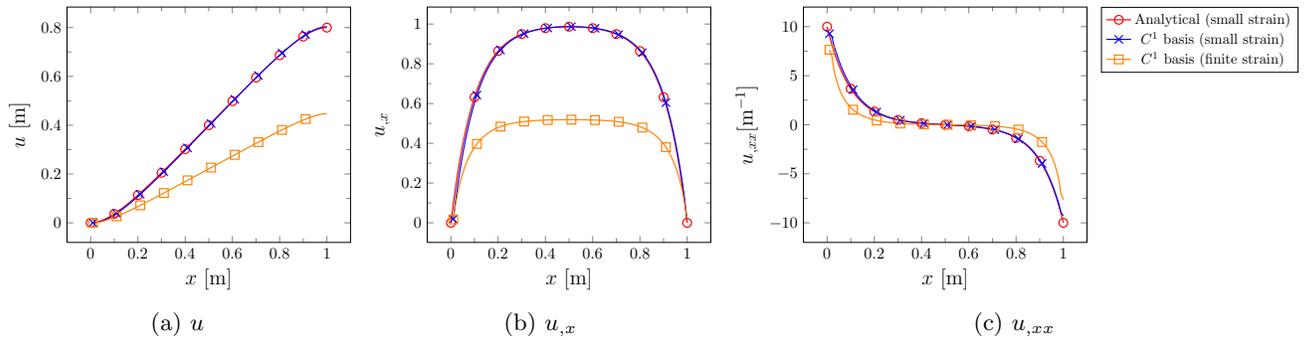
\subsection{Three-dimensional, uniaxial tension}
\label{sec:tension3}
\begin{figure}[h]
  \centering
  \psfrag{a}[][][0.75]{\small \begin{tabular}{c} (a) $Du_i=0\vert_{X_3=\{0,L\}}$ \\  $l=100.0$m, $|\bu|_{max}=0.008$m \end{tabular}}
  \psfrag{b}[][][0.75]{\small \begin{tabular}{c} (b) $Du_i=0\vert_{X_3=\{0,L\}}$ \\  $l=10.0$m, $|\bu|_{max}=0.613$m \end{tabular}}
  \psfrag{c}[][][0.75]{\small \begin{tabular}{c} (c) $l=0.0$m, $|\bu|_{max}=2.76$m \end{tabular}}
  \psfrag{x}[][][0.75]{\small 1m}
  \psfrag{y}[][][0.75]{\small 1m}
  \psfrag{z}[][][0.75]{\small L=10m}
  \psfrag{t}[][][0.75]{\small $1.0\text{Nm}^{-2}$}
  \psfrag{w}[][][0.75]{\small $u_3=0|_{X_3=0}$}
   \psfrag{k}[][][0.75]{\small $\be_1$}	
   \psfrag{l}[][][0.75]{\small $\be_2$}	
   \psfrag{m}[][][0.75]{\small $\be_3$}	
  \includegraphics[width=.6\textwidth]{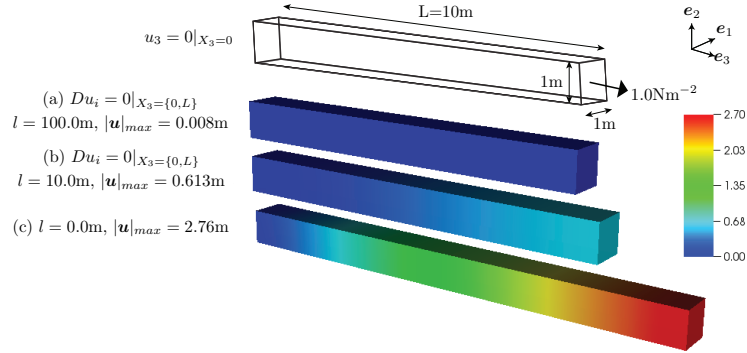}
  \caption{Effect of the gradient length scale parameter and the higher-order Dirichlet boundary condition on the deformation for the uniaxial tension boundary value problem. Contours show the displacement magnitude. Case (a) and (b) enforce $Du_i=0$ along the faces $X_3={0,L}$ and Case (c) is the result obtained for the non-gradient formulation ($l=0$).}
  \label{fig:3D:tensionUPlot}
\end{figure}

We consider uniaxial tension in three dimensions and study the effect of gradient elasticity on the deformation. The problem geometry, boundary conditions and the deformation are shown in Figure \ref{fig:3D:tensionUPlot}. The loading is a traction vector as specified. For all the three dimensional simulations in this paper, we use: $\lambda=1.0,\:\mu=1.0$. Furthermore, for each boundary value problem we vary the gradient length scale parameter $l$ to demonstrate the effect of gradient elasticity on the displacement field, strain energy and strain gradient energy distribution. \\

In the case of unixial tension, standard first order displacement or traction boundary conditions are not sufficient to induce strain gradients and hence the effect of increasing the gradient length scale parameter $l$ is negligible. However, when higher order Dirichlet boundary conditions  ($Du_i=0$) are enforced, strain gradients are induced at the boundaries and lead to a significant stiffening of the deformation response. Of course, on boundary faces where no higher order Dirichlet boundary condition is specified, the conjugate higher order Neumann boundary condition ($B_{iJK}N_J N_K = 0$) is implied. 

\begin{figure}[h]
  \centering
  \subfloat[\label{fig:tensionBCmaxU}]{
    \begin{tikzpicture}[scale=0.65]
      \begin{loglogaxis}[minor tick num=1,xlabel={\large $l$},ylabel={\large $|\bu|_{max}$}, x unit=m, y unit=m]
        \addplot [red, mark=x, mark size=4, thick] table [x={l}, y={Cubic_20_true_U}] {\tensionValues};
       	\addplot [orange, mark=square, mark size=2, thick] table [x={l}, y={Cubic_20_fals_U}] {\tensionValues};
        \legend{with $Du_i=0$, without $Du_i=0$}
      \end{loglogaxis}
    \end{tikzpicture}
  }
  \subfloat[\label{fig:tensionBCEnergy}]{
    \begin{tikzpicture}[scale=0.65]
      \begin{loglogaxis}[minor tick num=1,xlabel={\large $l$},ylabel={\large Energy}, x unit=m, y unit=J]
        	\addplot [red, mark=x, mark size=4, thick] table [x={l}, y={Cubic_20_true_SE}] {\tensionValues};
        \addplot [orange, mark=square, mark size=2, thick] table [x={l}, y={Cubic_20_true_GE}] {\tensionValues};
	\legend{Non-gradient energy, Gradient energy}
       \end{loglogaxis}
    \end{tikzpicture}
  }
  \caption{Effect of the gradient length scale parameter on the maximum displacement and elastic free energy contributions for the uniaxial tension boundary value problem. (a) Interaction of the higher order Dirichlet boundary condition and gradient length scale parameter on the maximum displacement value, and (b) strain energy and strain gradient energy contributions to the total elastic energy when the higher-order Dirichlet boundary condition is enforced.}
  \label{fig:tensionBCMaxUEnergy}
\end{figure}
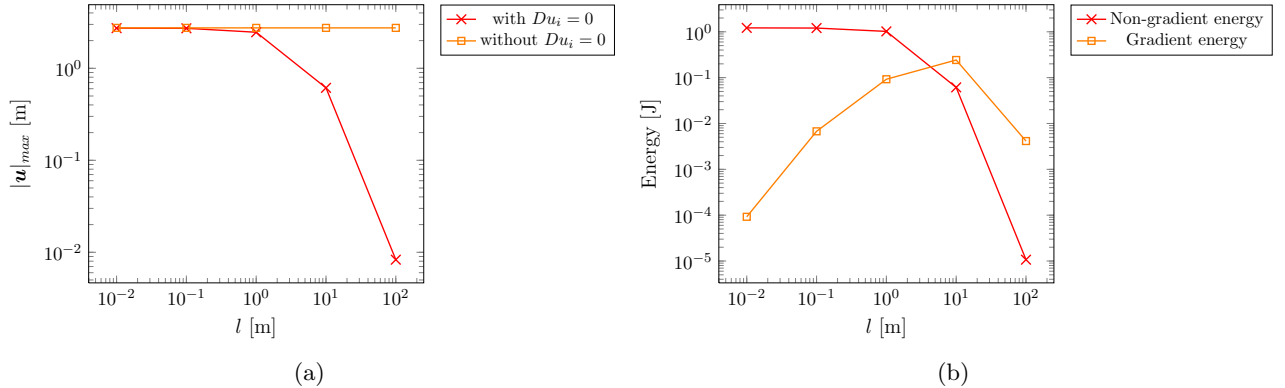

The interaction of the higher order Dirichlet boundary condition and gradient length scale parameter on the deformation response of the problem is shown in Figure \ref{fig:tensionBCMaxUEnergy}a. Furthermore under these conditions the energy distribution between the regular strain energy and strain gradient energy is shown in Figure \ref{fig:tensionBCMaxUEnergy}b. Clearly, for $l \ge 4$ m, the strain gradient energy becomes more significant than the regular strain energy, and further increase in $l$ leads to greater stiffening and thus insignificant deformation. \\ 

As shown in Figure \ref{fig:tensionBCMaxUEnergy}a, without the higher order boundary condition, strain gradient elastic effects are minimal in the uniaxial tension problem. This is the basis of our statement in the Introduction that the higher order boundary condition can dominate over the influence of the length scale parameter $l$ in inducing strain gradients in the solution. However, this changes in the case of bending and torsion, which we consider in the following sections. For these problems strain gradients are naturally induced by the standard boundary conditions and the effect of the higher order boundary condition is less significant.  

\subsection{Three-dimensional bending of a cantilever beam}
\label{sec:bending}
\begin{figure}[h]
  \centering
  \psfrag{a}[][][0.75]{\small \begin{tabular}{c} (a) $Du_i=0|_{X_3=\{0,L\}}$ \\  $l=10.0$m, $|\bu|_{max}=0.13$m \end{tabular}}
  \psfrag{b}[][][0.75]{\small \begin{tabular}{c} (b) $Du_i=0|_{X_3=\{0,L\}}$ \\  $l=1.0$m, $|\bu|_{max}=1.96$m \end{tabular}}
  \psfrag{c}[][][0.75]{\small \begin{tabular}{c} (c) $l=0.0$m, $|\bu|_{max}=8.17$m \end{tabular}}
  \psfrag{x}[][][0.75]{\small 1m}
  \psfrag{y}[][][0.75]{\small 1m}
  \psfrag{z}[][][0.75]{\small L=10m}
  \psfrag{t}[][][0.75]{\small $0.01\text{Nm}^{-2}$}
  \psfrag{w}[][][0.75]{\small $\bu=\boldmath{0}|_{X_3=0}$}
   \psfrag{k}[][][0.75]{\small $\be_1$}	
   \psfrag{l}[][][0.75]{\small $\be_2$}	
   \psfrag{m}[][][0.75]{\small $\be_3$}	
  \includegraphics[width=.5\textwidth]{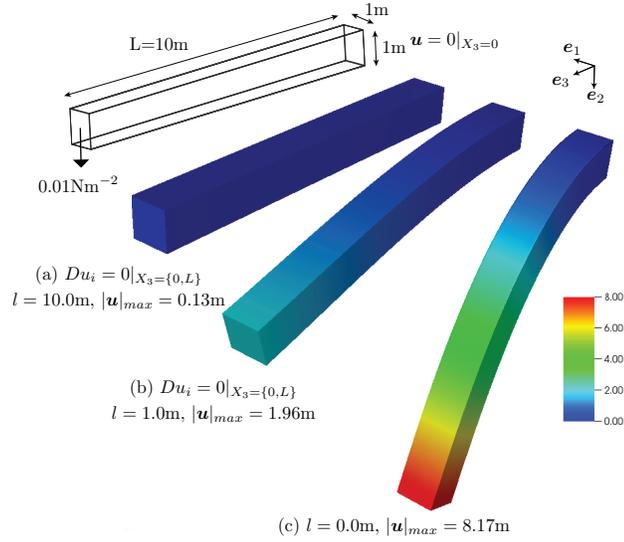}
  \caption{Effect of the gradient length scale parameter and the higher-order Dirichlet boundary condition on the deformation for the bending boundary value problem. Contours show displacement magnitude. Cases (a) and (b) enforce $Du_i=0$ along the faces $X_3=\{0,L\}$ and Case (c) is the result obtained for the non-gradient formulation ($l=0
 $).}
  \label{fig:3D:bendingUPlot}
\end{figure}

\begin{figure}[h]
  \centering
  \subfloat[\label{fig:bendingBCmaxU}]{
    \begin{tikzpicture}[scale=0.65]
      \begin{loglogaxis}[minor tick num=1,xlabel={\large $l$},ylabel={\large $|\bu|_{max}$}, x unit=m, y unit=m]
       	\addplot [red, mark=x, mark size=4, thick] table [x={l}, y={Cubic_20_true_U}] {\bendingValues};
        \addplot [orange, mark=square, mark size=2, thick] table [x={l}, y={Cubic_20_fals_U}] {\bendingValues};
        \legend{with $Du_i=0$, without $Du_i=0$}
      \end{loglogaxis}
    \end{tikzpicture}
  }
  \subfloat[\label{fig:bendingBCEnergy}]{
    \begin{tikzpicture}[scale=0.65]
      \begin{loglogaxis}[minor tick num=1,xlabel={\large $l$},ylabel={\large Energy}, x unit=m, y unit=J]
        \addplot [red, mark=x, mark size=4, thick] table [x={l}, y={Cubic_20_true_SE}] {\bendingValues};
        \addplot [orange, mark=square, mark size=2, thick] table [x={l}, y={Cubic_20_true_GE}] {\bendingValues};
        \legend{Non-gradient energy, Gradient energy}
      \end{loglogaxis}
    \end{tikzpicture}
  }
  \caption{Effect of the gradient length scale parameter on the deformation response and energy distribution for the bending boundary value problem. (a) Weak influence of the higher order Dirichlet boundary condition on the maximum deflection, and (b) strain energy and strain gradient energy contributions to the total elastic energy when the higher-order Dirichlet boundary condition is enforced.}
  \label{fig:bendingBCMaxUEnergy}
\end{figure}
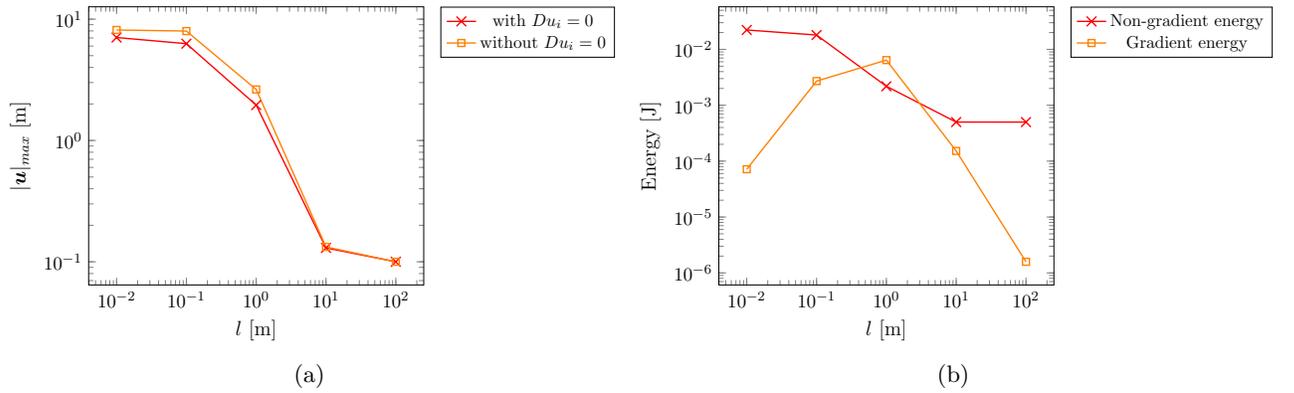

As can be appreciated from the Euler-Bernoulli theory for thin beams, the kinematics of bending induce strong strain gradients. In this section, we apply our framework to elucidate these effects. The same structure from Section \ref{sec:tension3} is now subjected to bending by specifying boundary conditions as shown in Figure \ref{fig:3D:bendingUPlot}. Note the loading by the traction vector. Figures \ref{fig:3D:bendingUPlot}a--\ref{fig:3D:bendingUPlot}c demonstrate the strong stiffening effect of strain gradient elasticity as the gradient length scale parameter increases. In this boundary value problem, the imposition of the higher-order Dirichlet boundary condition has a very weak influence in inducing strain gradient effects additional to those already present due to the kinematics of bending, as seen in Figure \ref{fig:bendingBCmaxU}. Figure \ref{fig:bendingBCEnergy} shows that the strain and strain gradient energy contributions differ by less than in uniaxial tension. This slightly tighter coupling is a result of the strong strain gradients present even in the non-gradient elasticity formulation of bending boundary value problems.
\subsection{Three-dimensional torsion of a cylinder with square cross section}
\label{sec:torsion3}

\begin{figure}[h]
  \centering
  \psfrag{a}[][][0.75]{\small \begin{tabular}{c} (a) $Du_i=0|_{X_3=\{0,L\}}$ \\  $l=10.0$ m, $|\bu|_{max}=0.001$ m \\ $u_3$ scale factor: $1.0\times 10^5$ \end{tabular}}
  \psfrag{b}[][][0.75]{\small \begin{tabular}{c} (b) $Du_i=0|_{X_3=\{0,L\}}$ \\  $l=1.0$ m, $|\bu|_{max}=0.099$ m  \\ $u_3$ scale factor: $1.0\times 10^3$ \end{tabular}}
  \psfrag{c}[][][0.75]{\small \begin{tabular}{c} (c) $l=0.0$ m, $|\bu|_{max}=0.538$ m  \\ $u_3$ scale factor: $1.0\times 10^1$ \end{tabular}}
   \psfrag{x}[][][0.75]{\small 1m}
  \psfrag{y}[][][0.75]{\small 1m}
  \psfrag{z}[][][0.75]{\small L=10m}
  \psfrag{t}[][][0.75]{\small $M_3=0.1\text{Nm}^{-1}$}
  \psfrag{w}[][][0.75]{\small $\bu=\boldmath{0}|_{X_3=0}$}
   \psfrag{k}[][][0.75]{\small $\be_1$}	
   \psfrag{l}[][][0.75]{\small $\be_2$}	
   \psfrag{m}[][][0.75]{\small $\be_3$}	
  \includegraphics[width=.65\textwidth]{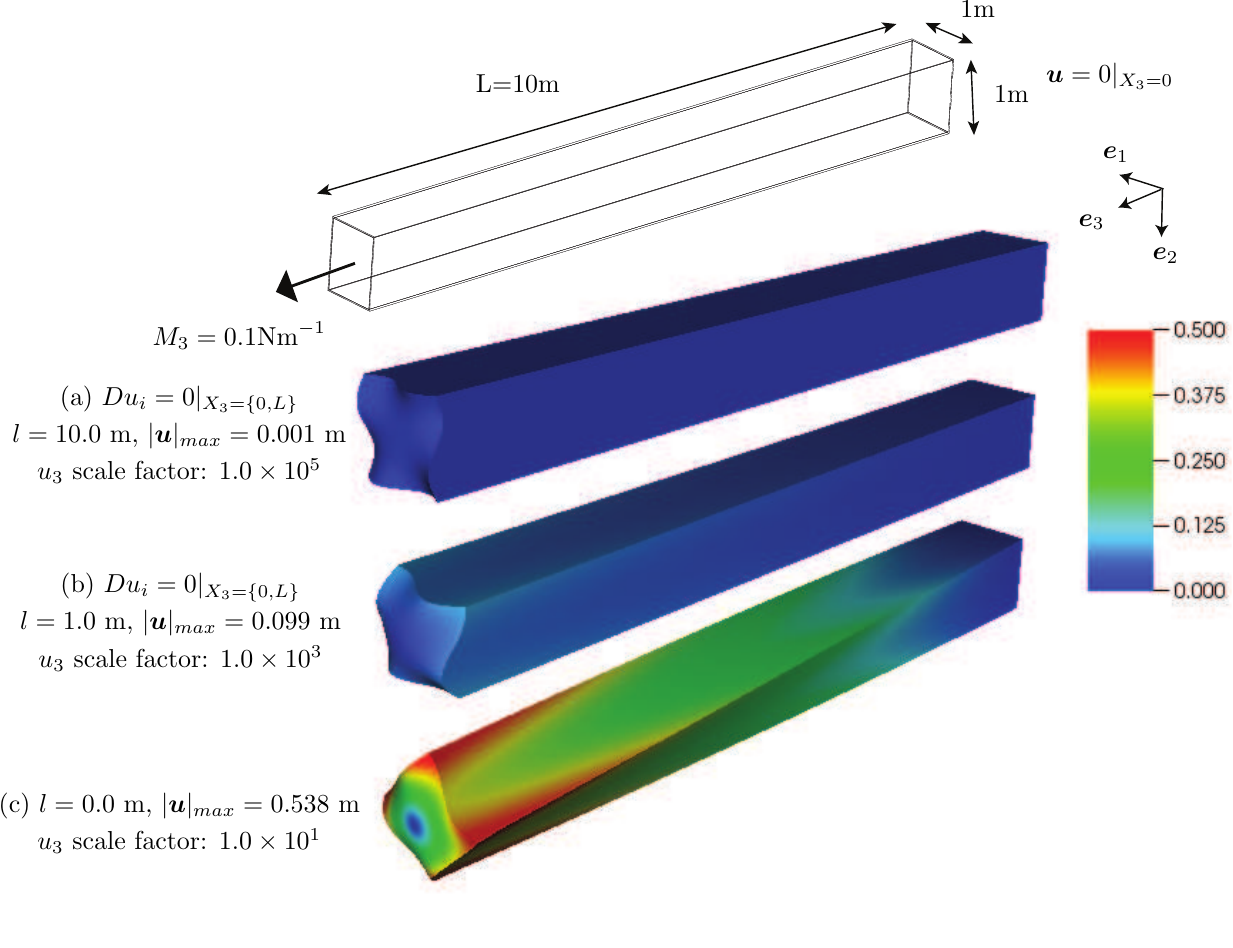}
  \caption{Effect of the gradient length scale parameter and the higher-order Dirichlet boundary condition on the deformation for the torsion boundary value problem. Contours show displacement magnitude. Case (a) and (b) enforce $Du_i=0$ along the faces $X_3=\{0,L\}$ and Case (c) is the result obtained for the non-gradient formulation ($l=0
 $). The displacement component $u_3$ has been scaled by the factor indicated to make the surface warping discernible.}
  \label{fig:3D:torsionUPlot}
\end{figure}

\begin{figure}[h]
  \centering
  \subfloat[\label{fig:torsionBCmaxU}]{
    \begin{tikzpicture}[scale=0.65]
      \begin{loglogaxis}[minor tick num=1,xlabel={\large $l$},ylabel={\large $|\bu|_{max}$}, x unit=m, y unit=m]
        \addplot [red, mark=x, mark size=4, thick] table [x={l}, y={Cubic_20_true_U}] {\torsionValues};
       	\addplot [orange, mark=square, mark size=2, thick] table [x={l}, y={Cubic_20_fals_U}] {\torsionValues};
        \legend{with $Du_i=0$, without $Du_i=0$}
      \end{loglogaxis}
    \end{tikzpicture}
  }
  \subfloat[\label{fig:torsionBCEnergy}]{
    \begin{tikzpicture}[scale=0.65]
      \begin{loglogaxis}[minor tick num=1,xlabel={\large $l$},ylabel={\large Energy}, x unit=m, y unit=J]
        \addplot [red, mark=x, mark size=4, thick] table [x={l}, y={Cubic_20_true_SE}] {\torsionValues};
        \addplot [orange, mark=square, mark size=2, thick] table [x={l}, y={Cubic_20_true_GE}] {\torsionValues};
         \legend{Non-gradient energy, Gradient energy}
      \end{loglogaxis}
    \end{tikzpicture}
  }
  \caption{Effect of the gradient length scale parameter on the deformation response and energy distribution for the torsion boundary value problem. (a) Virtually non-existent influence of higher order Dirichlet boundary condition on the maximum displacement value which reflects the angle of rotation, and (b) strain energy and strain gradient energy contributions to the total elastic energy when higher order Dirichlet boundary conditions are enforced.}
  \label{fig:torsionBCMaxUEnergy}
\end{figure}
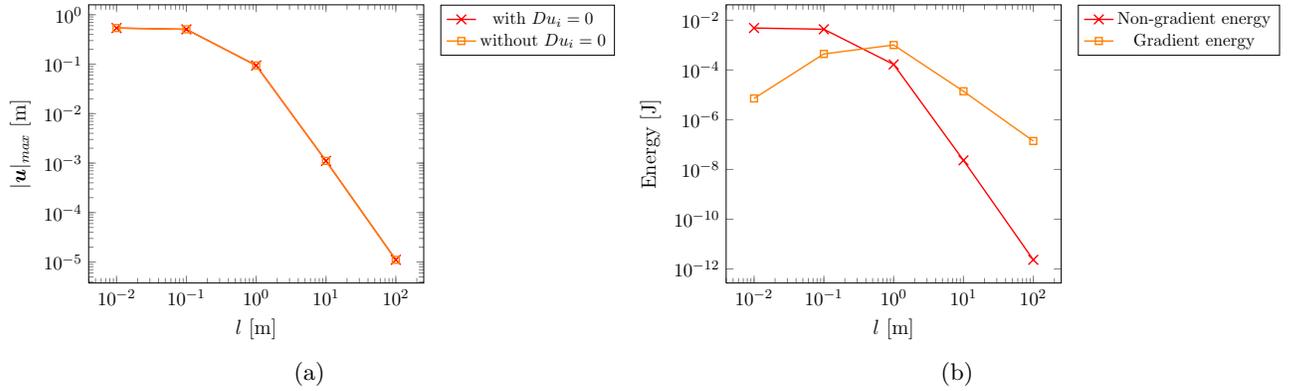

Like bending, the twisting kinematics induced by torsion throw up strong strain gradients even with the non-gradient elasticity formulation. We persist with the same structure; boundary conditions appear in Figure \ref{fig:3D:torsionUPlot}. Note that the loading in this case is due to an areal torque density vector as specified. The non-circular cross-section suffers warping as is well-known from the classical treatment of this problem. Figures \ref{fig:3D:torsionUPlot}a--\ref{fig:3D:torsionUPlot}c show the increased stiffness in torsion with an increase in gradient length scale parameter. This is apparent not only in the decrease in twist with increase in $l$, but also the fact that the warping displacement, $u_3$ has to be scaled up by two orders of magnitude between $l = 0$ m and $l = 1$ m, and again between  $l = 1$ m and $l = 10$ m, to be discernible on these plots. As in the case of bending, the strong strain gradients inherent in the kinematics of twisting even for the non-gradient elasticity formulation result in an almost non-existent influence of the higher-order Dirichlet boundary condition. This is seen in Figure \ref{fig:torsionBCmaxU}. Figure \ref{fig:torsionBCEnergy} shows the variation of the strain and strain gradient energy. Figures \ref{fig:tensionBCmaxU}, \ref{fig:tensionBCEnergy}; \ref{fig:bendingBCmaxU}, \ref{fig:bendingBCEnergy}; and \ref{fig:torsionBCmaxU}, \ref{fig:torsionBCEnergy} when considered together suggest that the difference between the strain energy and strain gradient energy contributions diminishes in problems wherein the kinematics induce strong strain gradients even in the non-gradient formulation.
%
%
%
\clearpage
\subsection{Discontinuities in higher-order stresses at edges}
\label{sec:line}
\begin{figure}[h]
  \centering
  \psfrag{a}[][][0.75]{\Large \begin{tabular}{c} (a) $l=0.0$m, $|\bu|_{max}=0.179$m \end{tabular}}
  \psfrag{b}[][][0.75]{\Large \begin{tabular}{c} (b) $l=0.1$m, $|\bu|_{max}=0.136$m  \end{tabular}}
  \psfrag{x}[][][0.75]{\small 1m}
  \psfrag{y}[][][0.75]{\small 1m}
  \psfrag{z}[][][0.75]{\small 1m}
  \psfrag{t}[][][0.75]{$L = 1.0\times 10^{-5} \be_3 ~\text{Nm}^{-1}$}
  \psfrag{w}[][][0.75]{\small $\bu=\boldmath{0}|_{X_3=0}$}
   \psfrag{k}[][][0.75]{\small $\be_1$}	
   \psfrag{l}[][][0.75]{\small $\be_2$}	
   \psfrag{m}[][][0.75]{\small $\be_3$}	
  \includegraphics[width=.8\textwidth]{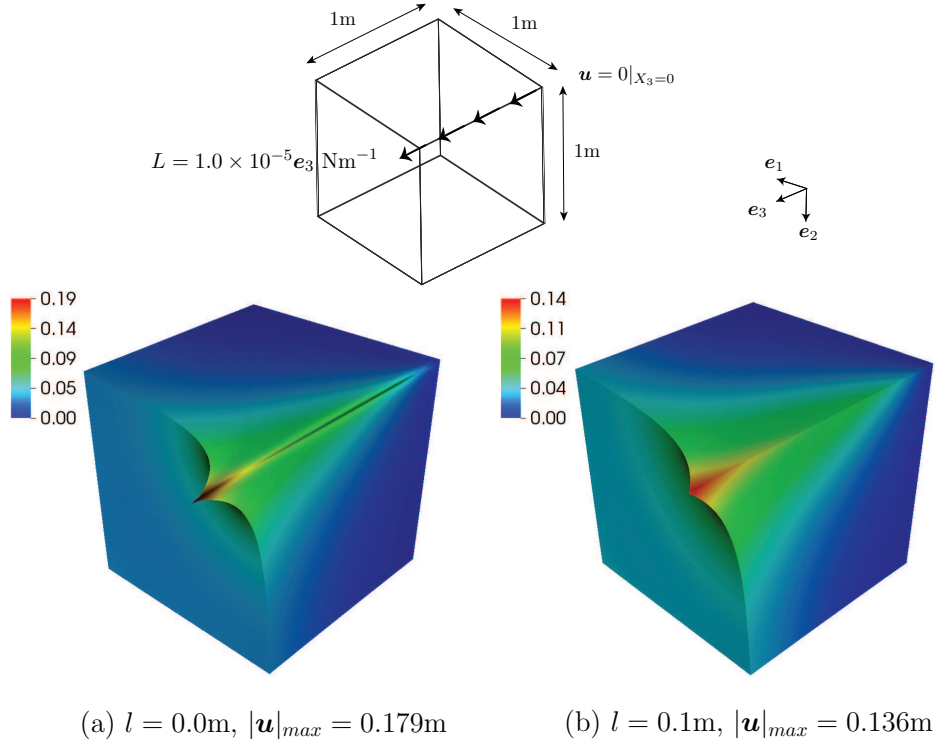}
  \caption{Effect of the length scale parameter on the deformation for the line traction boundary value problem. Case (a) is the non-gradient formulation where the line traction leads to a stress singularity and sharp displacement gradient. However, for Case (b) the gradient formulation naturally balances the line traction due to the higher order stress, such that $L_i = \llbracket B_{iJK}N_J^\Gamma N_K\rrbracket$. The resultant response is stiffer, as seen in the lower displacement under the line load.}
  \label{fig:3D:lineLoadUPlot}
\end{figure}

This boundary value problem demonstrates the effect of discontinuous higher-order stresses $B_{iJK}N_J^\Gamma N_K$, over smooth surfaces separated by an edge, $\Upsilon_{0^i}^L$. The discontinuity must be balanced by a line traction, $L_i = \llbracket B_{iJK}N_J^\Gamma N_K\rrbracket_{\Upsilon_{0^i}}$, as seen in Equation (\ref{eqn:strongformgradelasticity}). The resulting deformation appears in Figures \ref{fig:3D:lineLoadUPlot}a and \ref{fig:3D:lineLoadUPlot}b. Notably, whereas the non-gradient elasticity (Figure \ref{fig:3D:lineLoadUPlot}a) formulation yields a highly localized deformation in response to the line traction, the gradient elasticity formulation responds with a discontinuity in higher-order stress components: $\llbracket B_{iJK}N_J^\Gamma N_K\rrbracket_{\Upsilon_{0^i}} = L_i$, and a deformation response that gets stiffer as $l$ increases. 

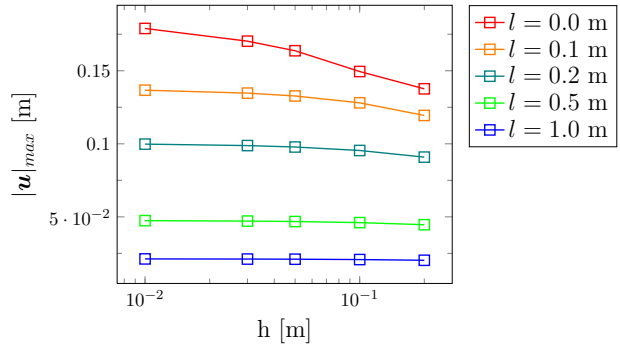
\begin{figure}[h]
  \centering
    \begin{tikzpicture}[scale=0.65]
      \begin{semilogxaxis}[minor tick num=1,xlabel={\Large h},ylabel={\Large $|\bu|_{max}$}, x unit=m, y unit=m]
        \addplot [red, mark=square, mark size=3, thick] table [x={h}, y={maxUval_l00}]{\lineLoad}; 
        \addplot [orange, mark=square, mark size=3, thick] table [x={h}, y={maxUval_l01}]{\lineLoad}; 
        	\addplot [teal, mark=square, mark size=3, thick] table [x={h}, y={maxUval_l02}]{\lineLoad}; 
        	\addplot [green, mark=square, mark size=3, thick] table [x={h}, y={maxUval_l05}]{\lineLoad}; 
         \addplot [blue, mark=square, mark size=3, thick] table [x={h}, y={maxUval_l1}]{\lineLoad}; 
         \legend{\Large {$l=0.0$ m}, \Large{$l=0.1$ m}, \Large{$l=0.2$ m}, \Large{$l=0.5$ m}, \Large{$l=1.0$ m}}
       \end{semilogxaxis}
    \end{tikzpicture}
  \caption{Effect of mesh refinement on the resolution of the maximum displacement for various values of the gradient length scale parameter. For $l=0.0$ m (the non-gradient formulation), the line traction leads to a stress singularity due to which the maximum displacement does not converge with respect to mesh refinement. However, for $l>0.0$ m, the gradient formulation eliminates the singularity, leading to convergence of the maximum displacement with mesh refinement.}
  \label{fig:lineLoad}
\end{figure}

It is well known that numerical solutions to non-gradient elasticity result in a stress singularity under a line load, which is a two-dimensional Dirac-delta function in $\mathbb{R}^3$. This manifests itself as a failure of the displacement field to converge with mesh refinement, and is seen in Figure \ref{fig:lineLoad} for gradient length scale parameter $l = 0$. In contrast, the higher-order character of the partial differential equation for gradient elasticity regularizes the solution, leading to a convergent solution with mesh refinement. See Figure \ref{fig:lineLoad} for $l > 0$. 

%

\section{Discussion}
\label{sec:discussion}
Our intent in this communication is to present our framework for the solution of general, three-dimensional boundary value problems with Toupin's theory. The numerical examples in Section \ref{sec:simulations} serve this purpose well because of the existing literature on linearized strain gradient elasticity solutions in one dimension (Section \ref{sec:1D}), and the familiarity with classical (non-gradient), nonlinear elasticity solutions to the remaining three-dimensional boundary value problems (Sections \ref{sec:tension3}--\ref{sec:line}). However, as we stated in the Introduction, strain gradient \emph{elasticity} is of physical relevance in the regime of sharp variation of the deformation at atomic length scales. This is almost never the case for the boundary value problems of Section \ref{sec:simulations}, which are typically applied at structural scales. In the Introduction, we did, however, identify two classes of problems for which Toupin's theory of gradient elasticity at finite strains is compelling: (a) if strains also serve as order parameters in representations of symmetry-lowering structural phase transformations, leading to a non-convex elastic free energy density, and (b)  if variations in deformation occur over length scales approaching inter-atomic distances, as at atomically sharp crack tips and dislocation cores. We now return to consider these problems.

\subsection{Strains as order parameters for structural transformations} 
\label{sec:martensitic}
An important class of structural phase transformations, commonly referred to as martensitic transformations, involve an affine deformation of the unit cell at the crystallographic level without any rearrangements of the atoms, such as by diffusional migration within the unit cell. The parent and final phases in such structural phase transformations are usually characterized by a symmetry group/subgroup relationship: A high symmetry phase can transform to several crystallographically equivalent lower symmetry variants. A classic example is the cubic to tetragonal  phase transformation involving an extension (contraction) of one of the cubic axes and a contraction (extension) of the remaining two cubic axes by an equal amount. Frame invariant strain metrics derived from the Cartesian components of $\bE$ can therefore also serve as order parameters tracking the degree and symmetry of the transformation product in this class of structural phase transformations. There are three symmetrically equivalent tetragonal variants that can emerge from the same cubic parent crystal. When the transformation strains are small, in the sense of a ``weak'' martensitic transformation as defined by \cite{Bhattacharyaetal2004}, the structural transformation can occur coherently, that is without introducing crystallographic defects such as dislocations. Typically, the transformation of a high symmetry phase into a lower symmetry phase results in a ``phase mixture'' of several symmetrically equivalent, lower-symmetry variants \citep{Ball1987, Ball1992, Ball2011}. The strain is very nearly uniform in each variant, suggesting the term ``lamina''.  The fine phase mixture of variants (laminae) can lead to a lower total elastic free energy, while maintaining compatibility, than if a single strained variant were to form. The mixture then consists of twin-boundaries separating the symmetrically equivalent lower symmetry variants. 

The existence of continuous strain order parameters that describe all phases participating in the structural transformation and that through variation deform one phase into another implies the existence of a continuous elastic free energy density as a function of strain. Such an elastic free energy density will then exhibit local minima for each of the mechanically stable low symmetry variants. If the high symmetry parent phase is an equilibrium structure, a smoothness requirement also applies to the elastic free energy density: The maxima separating the various local minima must be points of first-order differentiability, leading to an elastic free energy density that is smooth but \emph{non-convex} in strain space. The coherent coexistence of the various phases in a single microstructure will consist of boundaries where the strain order parameters vary continuously from one local free energy basin to another, crossing regions that are non-convex in highly (elastically) strained interface regions. Large gradients in strain will characterize the interface regions and the free energy of the microstructure will no-longer be accurately represented by an elastic free energy density that only depends on the local strain: non-local effects must be included in the free energy description. First-order contributions of non-local interactions can be captured with terms that depend on gradients of the strain metrics. 

\begin{figure}[h]
 \centering	
 \begin{tikzpicture}
\begin{axis}[y=2cm, no markers, width=0.8\textwidth, xlabel=$X$,ylabel=$u_1(X)$]
\addplot [red, thick] coordinates 
{(0,0) (0.5,0.5) (1,0)};
\addplot [orange, thick] coordinates 
{(0,0) (0.25,0.25) (0.5,0) (0.75,0.25) (1,0)};
\addplot [teal, thick] coordinates 
{(0,0) (0.125,0.125) (0.25,0) (0.375,0.125) (.5,0) (0.625,0.125) (0.75,0) (0.875,0.125) (1,0)};
\addplot [blue, thick] coordinates 
{(0,0) (0.0625, 0.0625) (0.125,0) (0.1875,0.0625) (.25,0) (0.3125,0.0625) (0.375,0) (0.4375,0.0625) (0.5,0) (0.5625, 0.0625) (0.625,0) (0.6875,0.0625) (.75,0) (0.8125,0.0625) (0.875,0) (0.9375,0.0625) (1.0,0)};
\end{axis}
\end{tikzpicture}
\caption{The sequence of minimizing displacement field solutions $u_1$, with the elastic free energy density $\widetilde{W}(\bE) = \frac{1}{4}E^4_{11} - \frac{1}{3}E^3_{11} - \frac{3}{4}E^2_{11}$, subject to the boundary conditions $u_1 = 0$ at $X_1 = \{0,1\}$, $u_2 = 0$ on $X_2 = 0$ and $u_3 = 0$ on $X_3 = 0$. The colors red, orange, green and blue are solutions for progressively finer phase mixtures.}
\label{fig:fineMinimizers}
\end{figure}
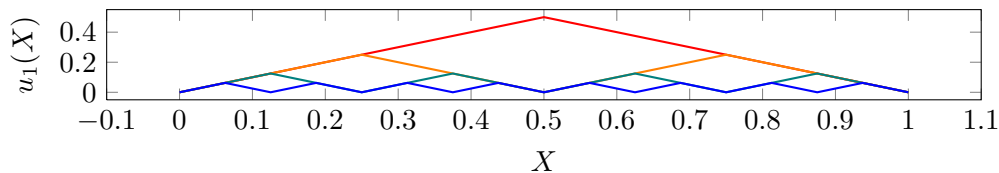

The non-convexity of the elastic free energy density with respect to particular strain metrics, in fact, requires the inclusion of strain gradient terms, not only to ensure a more accurate free energy description, but more fundamentally to also guarantee uniqueness of the displacement field that minimizes the total free energy. We illustrate this with a simple example. Let us consider an elastic free energy density function that is non-convex in strain space:
\begin{equation}
\widetilde{W}(\bE) = \frac{1}{4}E^4_{11} - \frac{1}{3}E^3_{11} - \frac{3}{4}E^2_{11}
\end{equation} 
\noindent over the domain $(0,1)\times (-b,b)\times (-c,c)$ subject to homogeneous Dirichlet boundary conditions $u_1 = 0$ at $X_1 = \{0,1\}$, $u_2 = 0$ on $X_2 = 0$ and $u_3 = 0$ on $X_3 = 0$. The solution to this boundary value problem is any sequence of laminae perpendicular to $\be_1$, with $\bF = \bF^\pm = \bone \pm \be_1\otimes\be_1$, almost everywhere, and uniform in each lamina. There are as many laminae with $\bF = \bF^+$ as with $\bF = \bF^-$. Any such microstructure is a minimizer of the elastic free energy, and there is an infinite sequence of such microstructures with increasingly finer laminae. Figure \ref{fig:fineMinimizers} is a plot of the displacement component $u_1$, for this sequence of microstructures. A lamina is delineated by each piecewise linear part of $u_1$, implying piecewise uniform deformation gradient $\bF^+$ or $\bF^-$, and represents one variant. Any Lipschitz function, $\bu(\bX)$, with $\bF = \bF^\pm$ almost everywhere and that attains the specified boundary conditions is a solution \citep{Muller1999}. The existence of an infinite sequence of solutions with an increasing fineness of mixture is a mark of non-uniqueness. It arises because the non-convex, multi-well, form of the elastic free energy density admits arbitrarily fine phase mixtures as solutions without penalizing the variation of the deformation gradient between $\bF^+$ and $\bF^-$. The non-uniqueness can be eliminated if the elastic free energy density is extended to also depend on gradients of $\bF$, thus regularizing the otherwise ill-posed problem of elasticity.

\noindent For $\widetilde{W}$ that is non-convex with respect to $\bE$, the extension to a dependence on $\textrm{Grad}\bE$ penalizes gradients in $\bE$, and therefore in $\bF$. Thereby, the discontinuity $\llbracket\bF\rrbracket = \bF^+ - \bF^-$ that develops at interfaces between laminae in our example above would be prevented. Instead, the free energy cost associated with $\textrm{Grad}\bE$ ensures that not all microstructures are global minimizers. In the example of Figure \ref{fig:fineMinimizers} the global minimizer is the solution with $\bF = \bF^+$ in $X_1\in (0,1/2-\delta)$,  $\bF = \bF^-$ in $X_1\in (1/2+\delta,1)$, or \emph{vice versa}, with $\delta$ being determined by the  strain gradient length scale parameter in $\widetilde{W}(\bE,\textrm{Grad}\bE)$. This is a natural regularization of the problem by  strain gradient elasticity \citep{Muller1999}. 

The use of strains as order parameters and the formulation of free energy densities that depend on local strain and strain gradients to describe martensitic phase transformations was formulated in large part by \cite{Barsch1984}. Weak structural transformations amenable to this thermodynamic description belong to a broader class of phase transformations where the participating phases share a symmetry group/subgroup relationship. \cite{vanderWaals1893} first introduced a gradient energy term when describing density fluctuations around a liquid-gas transformation. Similar gradient energy terms appear in Landau-Ginzberg free energies to describe fluctuations around second order phase transitions and in the \cite{CahnHilliard1958} and \cite{AllenCahn1972} treatments of spinodal decomposition and order-disorder reactions in alloys. The treatment of weak (coherent) structural transformations, however, has an added level of complexity when compared to other phase transformations described by a single continuous free energy surface as a function of relevant order parameters. The path and energetics of the structural transformation are highly sensitive to the actual microstructure due to the long-range elastic interactions among different phases and variants that affects the total free energy of the transforming material. Large transformation strains (i.e. strains connecting different free energy basins as a function of strains) force the use of finite strain gradient elasticity, which increases the nonlinearity very significantly.

\subsection{Strains at interatomic distances} 

Recall the treatment which opened the Introduction to this communication. If the structural dimensions are close to the atomic bond length, the deformation varies sharply over the bond length scale, and the elastic free energy density's dependence on gradients of $\bF$ becomes significant through Equation \eqref{eq:deformationMap}. In this regime size effects at the bond length scale become prominent, exemplified by the rapidly varying deformation at an atomically sharp crack tip, and the core of a dislocation. Formulations of strain gradient elasticity have been invoked to calculate singularity-free stress fields at crack tips \citep{Sternberg1967} and dislocation cores \citep{Lazaretal2006}. The relation between molecular models of solid mechanics and strain gradient elasticity at the nanoscale has been explored by \cite{Garikipati2003} and \cite{Maranganti2007}.

Simply expressing $W = \widehat{W}(\bF) = \widetilde{W}(\bE)$ does not suffice to parametrize the elastic free energy density, which again must be extended to $W = \widehat{W}(\bF,\textrm{Grad}\bF) = \widetilde{W}(\bE,\textrm{Grad}\bE)$ for frame invariance. A free energy penalty is incurred by sharp gradients of $\bE$ (or $\bF$), which prevents the development of unbounded solutions over finite domains, such as happens with classical elasticity at crack tips and dislocation cores. 

Strain gradient elasticity's regularization of the singular stress and strain fields predicted by classical elasticity at crack tips and dislocation cores has been appreciated previously, but the difficulty of obtaining analytic solutions has limited its applicability to them. Perhaps, now that it should be possible to compute solutions to these problems numerically, the corresponding gradient elasticity fields will begin making their appearance in semi-analytic solutions.

\subsection{Further applications of the framework}
We make the case that we have detailed a complete framework for the numerical solution of boundary value problems of finite strain gradient elasticity in three dimensions, and driven by boundary conditions that possess the full generality dictated by the theory. We also submit that, to the best of our knowledge, such a comprehensive solution framework has not been previously assembled for this problem. Our choice of Toupin's 1962 theory has been motivated by its generality, which itself is clearly an outcome of that author's rigorous wielding of variational tools. These origins have bestowed on Toupin's theory a complexity that has eluded general solutions until now. 

The search for numerical solutions has until recently been focused on surmounting the challenge posed by the fourth-order character of the strong form, or equivalently the necessity of working in the $\mathscr{H}^2$ function space in weak form. We have found isogeometric analytic methods to resolve this difficulty by their straightforward construction of $C^n$-continuous functions, which satisfy the requirement of the $\mathscr{H}^2$ function space. Of comparable importance for attainment of solutions with significant strain gradient fields ($\mathscr{H}^2$-norm) has been the imposition of higher-order boundary conditions, especially the higher-order Dirichlet condition, which we have weakly enforced. In this regard, while we have provided interpretations for the conjugate higher-order Neumann condition on the couple stress traction, we note that the line traction condition, while successfully enforced, remains in need of a satisfying physical interpretation in the three-dimensional setting. Analogies to the discontinuity of moments across edges on $C^0$ shells are helpful in this regard, but in want of generalization. Finally, while it remains possible that a correct linearization, ``by hand'', of the weak form can be achieved, our reliance on algorithmic differentiation has vastly simplified the implementation, while retaining algorithmic exactness. This third ingredient of our numerical framework attains indispensability for one particular class of problems as we explain next.

Even more compelling than the problems involving martenistic transformations may be the class of mechano-chemically driven solid-to-solid phase transformation problems in which the free energy density possesses non-convexities in strain and composition spaces. For this class of problems the composition itself serves as an order parameter and leads to the classical Cahn-Hilliard problem \citep{CahnHilliard1958} if restricted to composition space. As in the purely mechanical martensitic problem, the non-convexity of free energy density in strain space arises from the existence of multiple variants of the crystal structure of the newly transformed phase; however, this phase forms due to the composition-driven transformation from its parent phase. This crystal structure's symmetry point group is a sub-group of that of the parent phase. As discussed in Section \ref{sec:martensitic}, the mechanics sub-problem is governed by finite strain gradient elasticity. The solution of this coupled problem thus involves partial differential equations that are of fourth order in mechanics and chemistry, and every one of the three main ingredients of our numerical framework prove critical. In this case, our preliminary studies indicate that the numerical solution of this problem would remain completely intractable without algorithmic differentiation, in particular.

In addition to the above applications, a significant body of work has developed in the past two decades on strain gradient plasticity \citep{Fleck1997, Gao1999, Muhlhaus1991, DeBorst1992, Acharya2000, Gurtin2000, Gurtin2005}. While we have not addressed this class of problems here, our previous experience with them \citep{GarikipatiHughes2000, Regueiroetal2002, Garikipativm2003, Wellsetal2004, Molarietal2006, OstienGarikipati2008} suggests that the numerical framework developed here will be applicable to them.

We have attempted to justify our focus in this communication on the more classical boundary value problems of uniaxial tension (one and three dimensions), bending, torsion, and line loads (all in three dimensions), and to demonstrate the effect of gradient elasticity on these well-understood problems. For these problems, the extension of the elastic free energy density to strain gradients effectively penalizes rapidly varying strain fields, causing a stiffer overall response as has been amply demonstrated in the numerical section. The treatment of martensitic transformations, crack tip and dislocation core fields by Toupin's theory calls for a different focus and developments of much detail, which will be the subject of forthcoming work.
%
%
\section{Acknowledgement}
\label{sec:acknowledgement}

The mathematical formulation for this research has been carried out under an ongoing NSF CDI Type I grant: CHE1027729 ``Meta-Codes for Computational Kinetics''. The numerical formulation and computations have been carried out as part of research supported by the U.S. Department of Energy, Office of Basic Energy Sciences, Division of Materials Sciences and Engineering under Award \#~DE-SC0008637 that funds the PRedictive Integrated Structural Materials Science (PRISMS) Center at University of Michigan.

\section{Appendix}
\label{sec::appendix}
In this section, we present the derivation of the variational formulation in the current configuration. Similar to the reference configuration, in the current configuration we consider the boundary to be the union of a finite number of smooth surfaces $\Gamma$, smooth edges $\Upsilon$ and corners $\Xi$: $\partial \Omega = \Gamma \cup \Upsilon \cup \Xi$.  Furthermore spatial gradients with respect to the current configuration are denoted by lower case indices. For functions defined on $\partial \Omega$, when necessary, the gradient operator is decomposed into the normal gradient operator $D$ and the surface gradient operator $D_{k}$, 
\begin{align}
\psi_{,k} &= D \psi n_{k} + D_{k} \psi\nonumber\\
\textrm{where}\quad D \psi n_{k} &= \psi_{,i} n_{i}n_{k}\;\textrm{and}\; D_{k} \psi = \psi_{,k} - \psi_{,i} n_{i}n_{k}
\label{surfacenormalgradientCurrent}
\end{align}

\noindent The total elastic free energy of the system is given by the following functional defined over the current configuration:
\begin{equation}
	\Pi[\bu] = \int_{\Omega}  \widehat{W} (\bF, \mathrm{Grad}\bF)  ~\frac{\mathrm{d}v}{J} -  \int_{ \Gamma^T} \bu\cdot \bt \, ~\mathrm{d}s  - \int_{\Gamma^M} D \bu\cdot \bm\, ~\mathrm{d}s  - \int_{\Upsilon^L} \bu\cdot \bl \, \mathrm{d}c,
\label{eq:totalfreeenergyCurrent}
\end{equation}
where $J=det(\bF)$, $\bt$ is the surface traction, $\bm$ is the surface moment and $\bl$ is a line force. Following Equation \eqref{surfacenormalgradientCurrent}, $D \bu=(\partial \bu/ \partial \bx)\cdot\,\bn$ is the normal derivative of the displacement on the boundary. Furthermore, $\Gamma= \Gamma_{i}^u \cup  \Gamma_{i}^T = \Gamma_{i}^m \cup  \Gamma_{i}^M$ represents the decomposition of the smooth surfaces of the boundary and $\Upsilon= \Upsilon_{i}^l \cup ~\Upsilon_{i}^L$ represents the decomposition of the smooth edges of the boundary into Dirichlet boundaries (identified by superscripts $u, m ~\textrm{and} ~l$) and Neumann boundaries (identified by superscripts $T, M ~\textrm{and} ~L$). We are interested in a displacement field of the following form:
\begin{equation}
u_i \in \mathscr{S}, \;\textrm{such that}\; u_i = \bar{u}_i,\;\forall \bx \in \Gamma_{i}^u;\quad u_i = \bar{l}_i,\;\forall \bx\in\Upsilon_{i}^l;\quad Du_i = \bar{m}_i,\;\forall \bx\in\Gamma_{i}
\label{dirbcsuCurrent}
\end{equation}

At equilibrium, the first variation of the free energy with respect to the displacement field is zero. As is standard, to construct such a variation we first consider variations on the displacement field $\bu_\varepsilon := \bu + \varepsilon\bw$, where 
\begin{equation}
 w_i\in\mathscr{V}\;\textrm{such that}\;w_i = 0 ~\forall ~\bx \in \Gamma_{i}^u \cup  \Upsilon_{i}^l, ~Dw_i = 0 ~\forall ~\bx \in \Gamma_{i}^m 
\label{dirbcswCurrent}
\end{equation}
 We construct the first variation of the free energy with respect to the displacement. 
\begin{align}
  \frac{\delta}{\delta\bu}\Pi[\bu] = &~\frac{\mathrm{d}}{\mathrm{d} \varepsilon} \Pi[\bu_\varepsilon]
   \bigg|_{\varepsilon=0} \nonumber \\
  =  &\int_{\Omega} \left( \frac{\partial \widehat{W}}{\partial F_{iJ}} w_{iJ} +  \frac{\partial \widehat{W}}{\partial F_{iJ,K}}  w_{iJ,K} \right) ~\frac{\mathrm{d}v}{J}  \nonumber \\ 
  -  &\int_{\Gamma_{i}^T}  w_{i} t_{i} \, ~\mathrm{d}s  - \int_{\Gamma_{i}^M} Dw_i m_{i} \, ~\mathrm{d}s  - \int_{\Upsilon_{i}^L} w_{i} l_{i} \, \mathrm{d}c
\end{align}
But since,
\begin{align}
w_{i,J} &=w_{ij}~F_{jJ} \nonumber \\
w_{i,JK} &=\left( w_{i,j}~F_{jJ} \right)_{,K} = w_{i,jK}~F_{jJ}+w_{i,j}~F_{jJ,K} = w_{i,jk}~F_{kK}~F_{jJ}+w_{i,j}~F_{jJ,K} \nonumber 
\end{align}
we have,
\begin{align}
  \frac{\delta}{\delta\bu}\Pi[\bu]
  =  &\int_{\Omega} w_{i,j} \left( \frac{\partial \widehat{W}}{\partial F_{iJ}} F_{jJ} + \frac{\partial \widehat{W}}{\partial F_{iJ,K}} F_{jJ,K} \right) ~\frac{\mathrm{d}v}{J} +  
        \int_{\Omega} w_{i,jk} \left(\frac{\partial \widehat{W}}{\partial F_{iJ,K}} F_{kK} F_{jJ} \right) ~\frac{\mathrm{d}v}{J} \nonumber \\   
  -  &\int_{\Gamma_{i}^T}  w_{i} t_{i} \, ~\mathrm{d}s  - \int_{\Gamma_{i}^M} Dw_i m_{i} \, ~\mathrm{d}s  - \int_{\Upsilon_{i}^L} w_{i} l_{i} \, \mathrm{d}c
\end{align}
Here we denote, 
\begin{align}
\frac{1}{J}~\left( \frac{\partial \widehat{W}}{\partial F_{iJ}} ~F_{jJ} + \frac{\partial \widehat{W}}{\partial F_{iJ,K}} ~F_{jJ,K} \right) &= \sigma_{ij}\label{eqn:stressSigma} \\
\frac{1}{J}~\left( \frac{\partial \widehat{W}}{\partial F_{iJ,K}} ~F_{kK}~F_{jJ} \right) &=\beta_{ijk} \label{eqn:stressBeta}
\end{align}
where $\sigma_{ij}$ are the components of the non-classical cauchy stress tensor, and $\beta_{ijk}$ are the components of the higher-order stress tensor. So the weak form of mechanical equilibrium is given by:
\begin{equation}
  \int_{\Omega} \left( \sigma_{ij} w_{i,j} +  \beta_{ijk} w_{i,jk} \right) ~\mathrm{d}v - \int_{\Gamma_{i}^T} w_i t_i \, ~\mathrm{d}s  - \int_{\Gamma_{i}^M} Dw_i M_i \, ~\mathrm{d}s  - \int_{\Upsilon_{i}^L} w_i l_i \, \mathrm{d}c = 0
\label{eqn:weakformCurrent}
\end{equation}

\noindent Now we proceed to derive the strong form of the problem. Applying integration by parts to Equation \eqref{eqn:weakformCurrent} we obtain, 
\begin{align}
  &- \int_{\Omega} \sigma_{ij,j} w_{i} ~\mathrm{d}v - \int_{\Omega} \beta_{ijk,k} w_{i,j} ~\mathrm{d}v +  \int_{\Gamma} \sigma_{ij}w_{i}n_{j} ~\mathrm{d}s +  \int_{\Gamma} \beta_{ijk}w_{i,j}n_{k} ~\mathrm{d}s \nonumber \\
 &- \int_{\Gamma_{i}^T} w_i t_i \, ~\mathrm{d}s  - \int_{\Gamma_{i}^M} Dw_i m_i \, ~\mathrm{d}s  - \int_{\Upsilon_{i}^L} w_i l_i \, \mathrm{d}c = 0
\label{eqn:weakform2Current}
\end{align}
Applying integration by parts again, but only on the second volume integral, yields, 
\begin{align}
  &- \int_{\Omega} \sigma_{ij,j} w_{i} ~\mathrm{d}v + \int_{\Omega} \beta_{ijk,jk} w_{i} ~\mathrm{d}v - \underbrace{\int_{\Gamma} \beta_{ijk,k}w_{i}n_{j} ~\mathrm{d}s}_{\text{Integral A}} +  \int_{\Gamma} \sigma_{ij}w_{i}n_{j} ~\mathrm{d}s +  \underbrace{\int_{\Gamma} \beta_{ijk}w_{i,j}n_{k} ~\mathrm{d}s}_{\text{Integral B}} \nonumber \\
 &- \int_{\Gamma_{i}^T} w_i t_i \, ~\mathrm{d}s  - \int_{\Gamma_{i}^M} Dw_i m_i \, ~\mathrm{d}s  - \int_{\Upsilon_{i}^L} w_i l_i \, \mathrm{d}c = 0
\label{eqn:weakform3Current}
\end{align}
Expanding the term labelled as Integral A by repeated use of Equation \eqref{surfacenormalgradientCurrent},
\begin{align}
\int_{\Gamma} \beta_{ijk,k}w_{i}n_{j} ~\mathrm{d}s &=  \int_{\Gamma} \left( \beta_{ijk,l} \delta_{lk} \right) w_{i}n_{j} ~\mathrm{d}s \nonumber \\
 &=  \int_{\Gamma} \left( D \beta_{ijk} n_{l} + D_{l} \beta_{ijk} \right) \delta_{lk} n_{j} w_{i} ~\mathrm{d}s \nonumber \\
 &=  \int_{\Gamma} \left( D \beta_{ijk} n_{k} n_{j} +  D_{k} \beta_{ijk} n_{j} \right) w_{i} ~\mathrm{d}s.
\label{eqn:weakformACurrent}
\end{align}

\noindent Likewise expanding the term labelled as Integral B:
\begin{align}
\int_{\Gamma} \beta_{ijk}w_{i,j}n_{k} ~\mathrm{d}s &=  \int_{\Gamma} \left( D w_{i} n_{j} + D_{j} w_{i} \right) \beta_{ijk} n_{k} ~\mathrm{d}s \nonumber \\
 &=  \int_{\Gamma} D w_{i} \beta_{ijk} n_{j} n_{k} ~\mathrm{d}s  +   \underbrace{\int_{\Gamma} D_{j} w_{i} \beta_{ijk} n_{k} ~\mathrm{d}s}_{\text{Integral C}}.
\label{eqn:weakformBCurrent}
\end{align}

\noindent The term labelled as Integral C yields,
\begin{align}
\int_{\Gamma} D_{j} w_{i} \beta_{ijk} n_{k} ~\mathrm{d}s &= \int_{\Gamma} D_{j} \left( w_{i} \beta_{ijk} n_{k} \right) ~\mathrm{d}s  -  \int_{\Gamma} w_{i} D_{j} \left( \beta_{ijk} n_{k} \right) ~\mathrm{d}s \nonumber \\
&= \underbrace{\int_{\Gamma} D_{j} \left( w_{i} \beta_{ijk} n_{k} \right) ~\mathrm{d}s}_{\text{Integral D}}  -  \int_{\Gamma} w_{i} \left( D_{j} \left( B_{ijk} \right) n_{k} +  \beta_{ijk}D_{j}n_{k} \right) ~\mathrm{d}s.
\label{eqn:weakformCCurrent}
\end{align}
Using the integral identity $\int_{\Gamma} D_{i} f_{....} n_{j} ~\mathrm{d}s = \int_{\Gamma} (b^k_k n_i n_j - b_{ij}) f_{....} ~\mathrm{d}s + \int_{\Upsilon}  \llbracket n^{\Gamma}_{i} n_{j} f_{....} \rrbracket  ~\mathrm{d}l$ \citep{Toupin1962}, where $b_{ij}=-D_{i}n_j=-D_{j}n_i$ are components of the second fundamental form of the smooth parts of the boundary,  and $\bn^\Gamma = \Bxi\times\bn$, where $\Bxi$ is the unit tangent to the curve $\Upsilon$, Integral D gives,
\begin{equation}
\int_{\Gamma} D_{j} \left( w_{i} \beta_{ijk} n_{k} \right) ~\mathrm{d}s = \int_{\Gamma}  (b^l_l n_j n_k - b_{jk}) w_{i} \beta_{ijk} ~\mathrm{d}s + \int_{\Upsilon} \llbracket n^{\Gamma}_{j} n_{k} w_{i} \beta_{ijk} \rrbracket  ~\mathrm{d}l
\label{eqn:weakformDCurrent}
\end{equation}
Collecting terms from Equations (\ref{eqn:weakform3Current}--\ref{eqn:weakformDCurrent}),
\begin{align}
  -& \int_{\Omega} w_{i} \left(\sigma_{ij,j} - \beta_{ijk,jk} \right)~\mathrm{d}v \nonumber \\
  &+ \int_{\Gamma} w_{i} \left(\sigma_{ij}n_j - D\beta_{ijk}n_kn_j - 2D_j(\beta_{ijk})n_k - \beta_{ijk}D_jn_k + (b^l_ln_jn_k-b_{jk})\beta_{ijk} \right)~\mathrm{d}s \nonumber \\
  &+ \int_{\Gamma} Dw_i\beta_{ijk}n_jn_k ~\mathrm{d}s  \nonumber \\
  &+ \int_{\Upsilon} w_i \llbracket n^{\Gamma}_{j} n_{k} \beta_{ijk} \rrbracket  ~\mathrm{d}l \nonumber \\
  &- \int_{\Gamma_{i}^T} w_i t_i \, ~\mathrm{d}s  - \int_{\Gamma_{i}^M} Dw_i m_i \, ~\mathrm{d}s  - \int_{\Upsilon_{i}^L} w_i l_i \, \mathrm{d}c = 0
\label{eqn:weakformFinalCurrent}
\end{align}

Standard variational arguments, including the invocation of homogeneous boundary conditions on $w_i$ on $\Gamma_{i}^u\cup\Upsilon_{i}^l$ and $Dw_i$ on $\Gamma_{0}^m$, then lead to the strong form of mechanical equilibrium for a material of grade two in the current configuration: \\
\begin{equation}
\begin{array}{rcl}
\sigma_{ij,j} - \beta_{ijk,jk} &= 0 &\mathrm{in} ~\Omega\\
u_{i}  &= \bar{u}_i   &\mathrm{on} ~\Gamma_{i}^u\\
\sigma_{ij}n_j - D\beta_{ijk}n_kn_j - 2D_j(\beta_{ijk})n_k - \beta_{ijk}D_jn_k + (b^l_ln_jn_k-b_{jk})\beta_{ijk} &= t_{i} &\mathrm{on} ~\Gamma_{i}^T\\
Du_i  &= \bar{m}_i &\mathrm{on} ~\Gamma_{i}^m\\
\beta_{ijk}n_jn_k &= m_{i} &\mathrm{on}  ~\Gamma_{i}^M\\
u_{i}  &= \bar{l}_i  &\mathrm{on} ~\Upsilon_{i}^l\\
\llbracket n^{\Gamma}_{j} n_{k} \beta_{ijk} \rrbracket &= l_{i} &\text{\small on} ~\Upsilon_{i}^L\\ \\
 \qquad \Gamma= \Gamma_{i}^u \cup  \Gamma_{i}^T, \qquad \Gamma= \Gamma_{i}^m \cup  \Gamma_{i}^M, \qquad \Upsilon= \Upsilon_{i}^l \cup  \Upsilon_{i}^L&
\end{array}
\label{eqn:strongformgradelasticityCurrent}
\end{equation}

%
%
\clearpage
\bibliographystyle{elsart-harv}
\bibliography{references}
\end{document}